\documentclass[apj]{emulateapj}
\usepackage{epsfig}
\usepackage{mathptmx} 
\usepackage{graphicx}
\usepackage[breaklinks=true]{hyperref} 

\shorttitle{Galaxy Formation with Stars and MBHs}
\shortauthors{Kim et al.}

\begin{document}

\title{Galaxy Formation with Self-consistently Modeled Stars and Massive Black Holes. I: Feedback-regulated Star Formation and Black Hole Growth}

\author{Ji-hoon Kim \altaffilmark{1,2}}
\email{me@jihoonkim.org}
\author{John H. Wise \altaffilmark{3,4}} 
\author{Marcelo A. Alvarez  \altaffilmark{5}}
\author{Tom Abel \altaffilmark{1,2,6,7}}

\altaffiltext{1}{Kavli Institute for Particle Astrophysics and Cosmology, SLAC National Accelerator Laboratory, Menlo Park, CA, USA}
\altaffiltext{2}{Department of Physics, Stanford University, Stanford, CA, USA}
\altaffiltext{3}{Department of Astrophysical Sciences, Princeton University, Princeton, NJ, USA}
\altaffiltext{4}{Hubble Fellow}
\altaffiltext{5}{Canadian Institute for Theoretical Astrophysics, Toronto, ON, Canada}
\altaffiltext{6}{Zentrum f\"ur Astronomie der Universit\"at Heidelberg, Institut f\"ur Theoretische Astrophysik, Heidelberg, Germany}
\altaffiltext{7}{Heidelberg Institut f\"ur Theoretische Studien, Heidelberg, Germany}

\begin{abstract}
There is mounting evidence for the coevolution of galaxies and their embedded massive black holes (MBHs) in a hierarchical structure formation paradigm.  
To tackle the nonlinear processes of galaxy - MBH interaction, we describe a self-consistent numerical framework which incorporates both galaxies and MBHs.  
The high-resolution adaptive mesh refinement (AMR) code {\it Enzo} is modified to model the formation and feedback of molecular clouds at their characteristic scale of 15.2 pc and the accretion of gas onto a MBH.
Two major channels of MBH feedback, radiative feedback (X-ray photons followed through full 3D adaptive ray tracing) and mechanical feedback (bipolar jets resolved in high-resolution AMR), are employed.  
We investigate the coevolution of a $9.2\times10^{11} M_{\odot}$ galactic halo and its $10^{5} M_{\odot}$ embedded MBH at redshift 3 in a cosmological $\Lambda$CDM simulation. 
The MBH feedback heats the surrounding ISM up to $10^6$ K through photoionization and Compton heating and locally suppresses star formation in the galactic inner core.  
The feedback considerably changes the stellar distribution there.
This new channel of feedback from a slowly growing MBH is particularly interesting because it is only locally dominant, and does not require the heating of gas globally on the disk.
The MBH also self-regulates its growth by keeping the surrounding ISM hot for an extended period of time. 
\end{abstract}


\section{Introduction} \label{sec:1}

Ever since the discovery of the ubiquitous existence of supermassive black holes at the centers of massive galaxies \citep[e.g.][]{1995ARA&A..33..581K}, a plethora of evidence has accumulated to indicate the coevolution of galaxies and their embedded massive black holes (MBHs). 
The observed tight correlation between MBH masses and bulge velocity dispersions \citep{2000ApJ...539L...9F, 2000ApJ...539L..13G} have bolstered the idea that the fates of a host galaxy and its embedded MBH are fundamentally intertwined and heavily affected by each other's influence \citep{1998A&A...331L...1S, 2000MNRAS.311..576K, 2003ApJ...595..614W}.  

Recent observations provide more solid constraints on the coevolution of galaxies and MBHs.  
For example, cosmological star formation history and black hole accretion history are measured to be proportional to each other \citep[e.g.][]{2009ApJ...707.1566Z}.  
Merging of galaxies is believed to induce quasar activity \citep[e.g.][]{2008ApJS..175..356H}, and the existence of high-redshift quasars \citep{2006AJ....132..117F} indicate the rapid growth of black hole masses in the early phase of hierarchical structure formation, most likely by mergers \citep{2001ApJ...552..459H}.  
Unmistakably it is a complicated and highly nonlinear process for a galaxy to affect its embedded MBH, and vice versa.  
Therefore, developing a numerical tool which incorporates both galaxies and MBHs in one self-consistent framework is indispensible to fully comprehend their coevolution.  

The seminal work by \cite{2005MNRAS.361..776S} to include accretion and feedback of a MBH in a galactic simulation has been followed by many detailed investigations.  
These studies have helped extend our understanding of galaxy - MBH interaction in various contexts and scales:  
(a) Merging of Milky Way sized galaxies was simulated to show that quasar-like MBH feedback drives a massive gas outflow leading to quenched star formation, and to the observed $M_{\rm BH}-\sigma_{\rm bulge}$ relation \citep{2005ApJ...620L..79S, 2005MNRAS.361..776S, 2005Natur.433..604D, 2009ApJ...690..802J}.  
(b) Successive mergers of galaxies and MBHs were performed in a cosmological volume to yield a viable route to form high-redshift quasars \citep{2007ApJ...665..187L, 2009MNRAS.400..100S}.  
(c) MBH feedback at the center of a galaxy cluster was demonstrated to release sufficient energy to stop an overly cooled inflow of gas \citep{2007MNRAS.380..877S, 2009MNRAS.398...53B, 2010arXiv1003.4744T, 2010arXiv1004.1851D}.

Nonetheless, a comprehensive numerical understanding which incorporates both galaxies and MBHs is still missing, for various reasons.  
First and foremost, simulated galaxies do not match some of the most obvious aspects of observed galaxies.  
For example, simulated galaxies are prone to lock baryons into too many stars \citep[][and references therein]{2010MNRAS.404.1111G}, or contain bulge-dominated disks that are too centrally concentrated and have a greatly reduced angular momentum relative to those observed
\citep{2001MNRAS.327.1334V, 2007MNRAS.375...53K, 2009arXiv0909.4156P}. 
These problems are somewhat alleviated by lowering star formation efficiency and/or increasing stellar feedback \citep{2007MNRAS.374.1479G, 2009arXiv0909.4167P, 2011MNRAS.410.1391A}, or even by introducing a new powerful energy source such as MBH feedback.  
However, the former fix has not been entirely successful even with varied feedback parameters while the latter almost always powers large-scale gas outflow leaving behind a ``red and dead'' galaxy devoid of gas for a long time \citep{2004MNRAS.348.1078B, 2005ApJ...625..588K, 2005MNRAS.361..776S, 2010arXiv1003.4744T}.
Obviously numerical simulations are still missing one or more essential ingredients.  
It could be  the ignored physical processes such as stellar UV radiation and magnetic fields.  
Or it could be the inaccurate descriptions of MBH accretion and feedback. 

Second, most numerical studies to date lack necessary resolution and technique to describe how gas falls onto a central MBH and how the energy input of MBH feedback is deposited to its surrounding gas.  
While the 1 - 100 kpc resolution in large-scale simulations is clearly insufficient to adequately describe the accretion flow onto a MBH, even galactic scale simulations do not generally resolve the Bondi radius (See  \S \ref{sec:2-accretion}; Eq.(\ref{eq:Bondi})), which is required in order to  trace how a MBH gravitationally influences its surroundings and how the radiation and outflows from the MBH are thermally coupled to the gas.  
Indeed, poor resolution has forced simulators to skip the thermalization process below the resolution limit, and to simply thermodynamically deposit MBH feedback energy near the MBH.  
While crude, it has been an effective approximation characterizing MBH feedback on a resolved scale  \citep{2005MNRAS.361..776S}.
And it might be a fairly reasonable choice if MBH feedback is powerful enough to drive thermal shock waves (so-called ``quasar-mode''; $\dot M_{\rm BH} > 0.02 \,\dot M_{\rm Edd}$).  
However, it can not adequately describe the energy coupling of the radiation from a weak, quiescent MBH \citep[``radio-mode''; $\dot M_{\rm BH} < 0.02 \,\dot M_{\rm Edd}$;][]{2006MNRAS.365...11C, 2007ARA&A..45..117M}.
For this reason injecting thermal energy in a small volume of poorly resolved interstellar medium (ISM) can hardly be an accurate description of MBH feedback (See \S \ref{sec:2-accretion} for detailed discussion).  
Modeling how MBH feedback energy is {\it actually} coupled to the gas is a critical missing piece in contemporary galaxy formation simulations.

Third, partly due to the lack of proper resolution, most numerical calculations to date have modeled stars and MBHs with phenomenologically parametrized {\it ad hoc} formulations.  
Most notably, the Eddington-limited Bondi-Hoyle accretion estimate    
 employed by many authors \citep[e.g.][See \S \ref{sec:2-accretion} for definitions of variables]{2005MNRAS.361..776S, 2005Natur.433..604D, 2009ApJ...690..802J} has had to be empirically boosted by an efficiency parameter $\alpha=$10 - 300.
\begin{eqnarray}
\dot M_{\rm BH} &=& {\tt min} \left( {4 \pi \alpha G^2 M^2_{\rm BH} \rho_{\rm B}  \over  { c^3_{\rm s} }    } \,\,,\,\,  {4 \pi G M_{\rm BH} m_{\rm p}  \over  { \epsilon_{\rm r} \sigma_{\rm T} c } }  \right).
\label{eq:Springel}
\end{eqnarray}
While this nondimensional boost factor $\alpha$ is to correct the large-scale averaged, and probably underestimated $\rho_{\rm B}$ near the MBH, $\alpha$ is typically fixed after the MBH has grown so the Bondi radius is resolved even with coarse resolution.\footnote{For more discussion on the boost factor $\alpha$, see \cite{2009ApJ...690..802J} or \cite{2009MNRAS.398...53B}.  For a discussion on the accretion rate adopted in this work, see \S \ref{sec:2-accretion} and Appendix \ref{sec:appendix-A}.}    
Another example of introducing tunable parameters based on unknown physics is to use two different implementations of MBH feedback, depending on the estimated accretion rate: quasar-mode feedback and radio-mode feedback \citep{2007MNRAS.380..877S, 2008ApJ...687L..53P}.  
While useful in some applications, these {\it ad hoc} approaches ironically demonstrate that the physics of MBHs has not yet been adequately described in simulations. 

In order to circumvent the limitations of previous approaches outlined above and to follow the actual physical processes between gas, stars, and MBHs, we develop a {\it fully self-consistent} galaxy formation simulation integrating the growths of both galaxies and MBHs in one comprehensive framework.   
We limit the use of {\it ad hoc} formulation but instead more accurately model the physics in all aspects of galaxy formation, namely: (a) molecular cloud formation, (b) stellar feedback, (c) MBH accretion, and (d) MBH feedback. 
Our code models the formation and feedback of molecular clouds at their characteristic scale of 15.2 pc (\S \ref{sec:2-formation} to \ref{sec:2-SF}) and the accretion of gas onto a MBH (\S \ref{sec:2-accretion}).  
Two major channels of MBH feedback are also considered: radiative feedback (monochromatic X-ray photons followed through full three dimensional adaptive ray tracing; \S \ref{sec:2-RF}) and mechanical feedback (bipolar jets resolved in high-resolution adaptive mesh; \S \ref{sec:2-MF}).   
We then investigate the evolution of a $9.2\times10^{11} M_{\odot}$ galactic halo with an embedded seed MBH of $10^{5} M_{\odot}$ at $z\sim3$ in a cosmological $\Lambda$CDM simulation. 

This paper will be the first in a series that assembles a number of high-resolution galaxy formation simulations with self-consistently modeled stars and MBHs.  
This article is organized as follows.  
The physics of galaxy formation in our code is the topic of \S \ref{sec:2}, followed by the initial condition of our simulation in \S \ref{sec:3}.  
\S \ref{sec:4} is devoted to the results of our experiments, with an emphasis on the feedback-regulated star formation and black hole growth.
Discussed in \S \ref{sec:5} are the conclusions and the future work.  

\section{Modeling the Physics of Galaxy Formation}\label{sec:2}

The high-resolution Eulerian adaptive mesh refinement (AMR) code {\it Enzo-2.0} \citep[http://enzo.googlecode.com/;][]{1997ASPC..123..363B, 1999ASSL..240...19N, 2001astro.ph.12089B, 2004astro.ph..3044O, 2007arXiv0705.1556N} captures the gravitational collapse of turbulent fragmentation with high spatial resolution \citep{2007ApJ...665..899W, 2008ApJ...682..745W, 2009Sci...325..601T} and attains multiphase gas dynamics in the ISM as it sharply resolves shocks and phase boundaries \citep{2005MNRAS.356..737S, 2007MNRAS.380..963A, 2008MNRAS.390.1267T}.  
Our enhanced version of {\it Enzo} contains all relevant features previously discussed in simulating galaxy evolution \citep{2006ApJ...641..878T, 2008ApJ...673..810T, 2008AIPC..990..429K, 2009ApJ...694L.123K} as well as a treatment of several new physical processes discussed in detail in the following sections.

\subsection{Hydrodynamics and Gravitational Dynamics}\label{sec:2-hyd}

The ZEUS astrophysical hydrodynamics module included in {\it Enzo}  is employed to solve the Euler equations for the collisional baryon fluid represented by grids \citep{1992ApJS...80..753S, 1992ApJS...80..791S, 1994ApJ...429..434A}.  
While known to introduce spurious effects, this scheme is widely used with AMR because of the stability of its solutions, and the acceptable error when combined with high resolution.

Dark matter, stars, and MBHs are treated as collisionless particles which interact only by the gravitational force.  
To evolve the particle positions and velocities, the gravitational dynamics are solved by an N-body adaptive particle-mesh solver.  
After particles are gridded onto the mesh by the cloud-in-cell interpolation, the Poisson equation is solved on the discretized density grids via fast Fourier transform and multigrid solvers \citep{1988csup.book.....H, 2004astro.ph..3044O}.

\subsection{Refinement Strategy}\label{sec:2-ref}

{\it Enzo} decides whether each parent cell needs to be refined into eight child cells based on the mass of the cell in gas or in particles.  
The timestep is also adaptively determined level by level so that the timestep $dt$ satisfies 
\begin{eqnarray}
dt &\leq 0.3 \times { \Delta x \over c_{\rm s} } = 0.3 \times {\rm (sound\,\,crossing \,\, time)}
\end{eqnarray}
for all the cells at that level.  
Here $c_{\rm s}$ is the sound speed of the gas, and we choose the Courant-Friedrichs-Lewy (CFL) safety number of 0.3.  
As decisions for refinement are made recursively, the resulting dataset is a nested grid-patch structure.  In our work, the grids are adaptively refined down to 15.2 pc resolution.  
This value is in accord with the Jeans length for a dense gas clump of $n=125$ ${\rm cm^{-3}}$ at $\sim$200 K, at which point a corresponding Jeans mass of  $16000 \,\,M_{\odot}$ collapses to spawn a molecular cloud particle.  

We refine the cells by factors of 2 in each axis, on gas and particle overdensities of 8.  
The mass thresholds, $M_{\rm ref}$, above which a cell refines are functions of a refinement level {\it l} as
\begin{eqnarray}
M_{\rm ref, gas}^{l} &=& 2^{-0.378 l} M_{\rm ref, gas}^0 = 2^{-0.378 l} \cdot 0.125\Omega_{\rm b} \rho_0 \Delta x^3\\
M_{\rm ref, part}^{l} &=& 2^{-0.105 l} M_{\rm ref, part}^0 = 2^{-0.105 l} \cdot 0.125 \Omega_{\rm m} \rho_0 \Delta x^3
\end{eqnarray}
where factors $0.125 = 8(1/2^3)^2$ guarantees to refine all the cells of the first two nested levels (\S \ref{sec:3-setup}).  
$\Delta x$ is the cell size at a root grid, and $\rho_0 = 3 H^2 / 8 \pi G$ is the critical density.  $\Omega_{\rm m} = 0.27$, $\Omega_{\rm b} = 0.044$, and $H = 71 {\, \rm km\,\, s}^{-1} {\rm Mpc}^{-1}$ are matter density, baryon density, and the Hubble constant, respectively.

For example, at the finest static level, $l = 2$, a cell is refined if it has more mass than $8.9\times10^5 M_{\odot}$ in gas or $6.7\times10^6 M_{\odot}$ in particles.  
At level $l = 11$ ($\Delta x =15.2 \,\,{\rm pc}$ at $z=3$) a cell is refined if more than $8.4\times10^4 \,\,M_{\odot} = 5 \,\, M_{\rm Jeans} (125\,\,{\rm cm^{-3}}, 200\,\,{\rm K})$ in gas or $3.5 \times 10^6 \,\,M_{\odot} = 47 \,\, M_{\rm DM, smallest}$ in particles.
This way we refine the grids more on small scales, which allows us to focus our computational resources more on the dense star forming regions, making the simulation super-Lagrangian \citep{2008ApJ...673...14O}.

\subsection{Chemistry and Radiative Cooling}

We use non-equilibrium chemistry model to track six species (H, ${\rm H}^+$, He, ${\rm He}^+$, ${\rm He}^{++}$, ${\rm e}^-$) by following six collisional processes among them.  
At the same time, {\it Enzo}'s cooling module considers collisional excitation cooling, collisional ionization cooling, recombination cooling, Bremsstrahlung cooling, and CMB Compton cooling to compute the radiative loss of internal gas energy \citep{1997NewA....2..209A, 1997NewA....2..181A}.  
Added to these primordial cooling rates is the metallicity-dependent metal cooling rate $\Delta \Lambda(Z) = \Lambda_{\rm net}(Z) - \Lambda_{\rm net}(0)$  above $10^4$ K, where $\Lambda_{\rm net}$ is the net cooling rate tabulated in \cite{1993ApJS...88..253S}.  
Cooling below $10^4$ K is also enabled with fine structure metal-line cooling by C, O, and Si \citep{2007ApJ...666....1G, 2008ApJ...685...40W}.  
This treatment ensures that a thin galactic disk forms by being cooled below $10^4$ K, the approximate virial temperature of the ISM in a galactic disk.  

We further refine our module with photoionization heating at $z < 3$ by the metagalactic background UV of quasars and galaxies \citep{1996ApJ...461...20H, 2001cghr.confE..64H}, which is known to give rise to a  warm diffuse ISM and prevent star formation in optically thin gas \citep{2009ApJ...695..292C}.  
An approximate self-shielding factor is applied when the heating term is added \citep{2005ApJ...635...86C}.  
While not introducing a marked difference in overall results analyzed here, inclusion of this additional heating term results in a more realistic interstellar medium.

\begin{figure}[t]
\epsscale{1.17}
\plotone{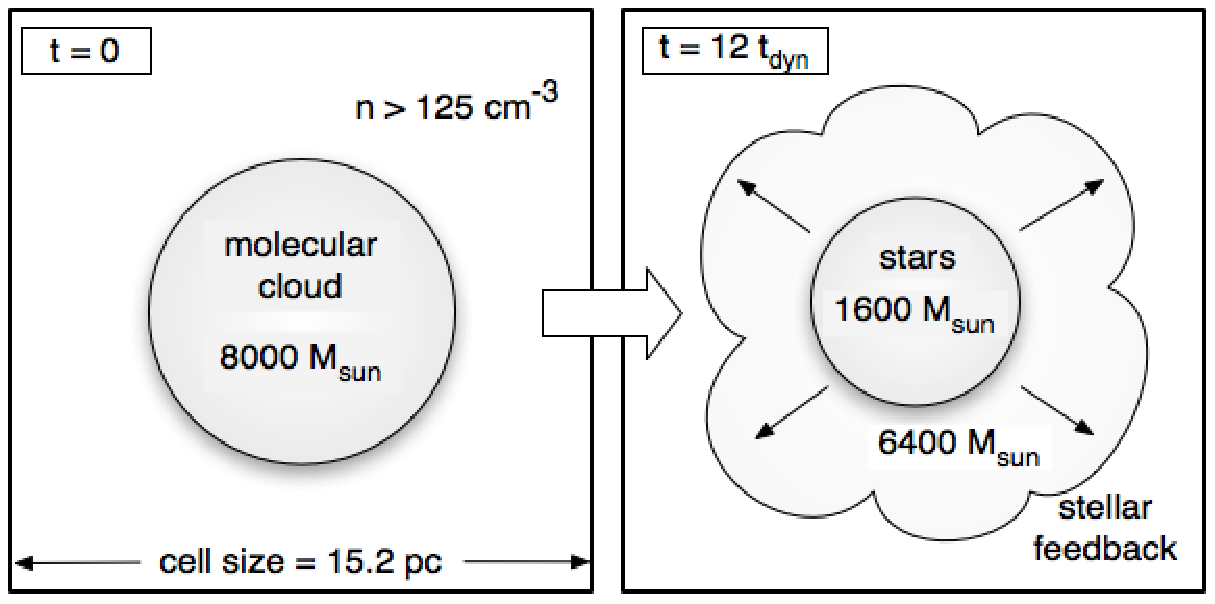}
\caption{A schematic view of the molecular cloud formation and the stellar feedback.  $12\,t_{\rm dyn}$ after a molecular cloud particle ($M_{\rm MC} \simeq 8000 \,\,M_{\odot}$) is formed, only 20\% of its mass remains as an actual stellar mass $M_{\rm star}(t)$ while the rest 80\% has returned to the gas along with thermal feedback energy.
\label{fig:SF}}
\end{figure}

\subsection{Molecular Cloud Formation}\label{sec:2-formation}

Our {\it molecular cloud} particle formation is based on \cite{1992ApJ...399L.113C} formalism with several important modifications.  
With a fixed formation efficiency of $\epsilon_* = 0.5$, the finest cell of physical size $\Delta x = 15.2 \,\,{\rm pc}$ and gas density $\rho_{\rm gas}$  produces a molecular cloud particle of mass
\begin{eqnarray}
M_{\rm MC} = \epsilon_* \rho_{\rm gas} \Delta x^3
\end{eqnarray}
when the following four criteria are met:
\begin{itemize}
\item[(a)] the proton number density exceeds the threshold $n_{\rm thres}=125$ ${\rm cm^{-3}}$,
\item[(b)] the velocity flow is converging; i.e. $\nabla \cdot \bold v < 0$,
\item[(c)] the cooling time $t_{\rm cool}$ is shorter than the dynamical time $t_{\rm dyn}$ of the cell: $E_{\rm int} / \dot{E} < [3\pi / (32G\rho_{\rm gas})]^{1/2}$, and
\item[(d)] the particle produced has at least $M_{\rm thres}  = 8000 \,\,M_{\odot}$.
\end{itemize} 
The consequence of our criteria is the following. 
The gas in the finest cell is converted into a particle as soon as the cell has accumulated more than $M_{\rm thres} / \epsilon_*= 16000 \,\,M_{\odot}$, the Jeans mass at $n = 125$ ${\rm cm^{-3}}$ at $\sim$200 K.   
Because 8000 - 42000 $M_{\odot}$ is instantly removed from the cell every time a particle is created, the gas mass in the finest cell never reaches the refinement threshold $M_{\rm ref, gas}^{l=11} = 84000\,\, M_{\odot}$ described in \S \ref{sec:2-ref}, ensuring the consistency between the refinement criteria and the particle formation.\footnote{Readers should be cautioned that the mass resolution of the reported simulation is $84000\,\, M_{\odot} = 5 \,\, M_{\rm Jeans} (125\,\,{\rm cm^{-3}}, 200\,\,{\rm K})$.  Ideally, if one properly combines the refinement strategy and the molecular formation criteria, the local Jeans mass can be resolved this way without explicitly requiring it.}
The values used here are in good agreement with those corresponding to collapsing Giant Molecular Clouds \citep[GMC;][]{2007ARA&A..45..565M} where star-forming molecular clumps are enshrouded by cold atomic gas.  

As an additional note, differences from more traditional star particle formation criteria such as in \cite{2008ApJ...673..810T} include: 
(a) the Jeans condition $\rho_{\rm gas} \Delta x^3 > M_{\rm Jeans}$ is removed because this condition could have allowed mass greater than $M_{\rm Jeans}$ to accumulate while not being properly resolved until a particle finally forms,
(b) the factor $\Delta t / t_{\rm dyn}$ in Eq.(1) of  \cite{2008ApJ...673..810T} is removed in order to instantly create a particle and not  leave any unresolved mass behind, and
(c) stochastic star formation is not imposed.  

With these modifications, our criteria guarantee that a particle forms before an unphysically large mass begins to accrete onto any unresolved dense clump.  
It is worth to emphasize the differences between our molecular cloud formation criteria and prior studies.   
While many previous studies with particle-based codes \citep[e.g.][]{1991ApJ...377...72M, 1992ApJ...391..502K, 1994ApJ...437..611M, 2003MNRAS.339..289S,  2007MNRAS.374.1479G} place a star particle using the Schmidt relation \citep[$\rho_{\rm SFR} \sim \rho_{\rm gas}^{1.5}$;][]{1959ApJ...129..243S}, we deposit a particle when a gas cell of a typical molecular cloud size actually becomes Jeans unstable.
For this reason, the particle in our simulation represents a {\it star-forming molecular cloud} that is self-gravitating, is thus decoupled from the gas on the grid.\footnote{We point out that the usual terminology of {\it star particle} to represent $10^5 - 10^6 M_{\odot}$ has been a misnomer.  We therefore make each of our particles to be 8000 $M_{\odot}$, regarding it as a molecular cloud gradually spawning stellar mass in it.  These particles are still collisionless and do not fully represent the real nature of molecular clouds.  However, we emphasize that our molecular cloud particles harbor a slow star formation rate matching observations.}
It is tagged with its mass $M_{\rm MC}$, dynamical time $t_{\rm dyn} = {\tt max} ([3\pi / (32G\rho_{\rm gas})]^{1/2}, 1.0 \,\,{\rm Myr})$, creation time $t_{\rm cr}$, and metallicity.   
Each molecular cloud particle gradually yields an {\it actual} stellar mass,  $M_{\rm star}(t)$, over 12 $t_{\rm dyn}$ which then contributes to stellar feedback (See Figure \ref{fig:SF} and \S \ref{sec:2-SF}). 

\subsection{Stellar Feedback}\label{sec:2-SF}

Observational evidence suggests that only $\sim$ 2\% of the gas in GMCs is converted into stars per dynamical time \citep[][and references therein]{2005ApJ...630..250K, 2007ApJ...654..304K}.  
Numerical studies also indicate that turbulence, magnetic fields, or radiation pressure can make the star formation process surprisingly slow \citep[e.g.][]{2010ApJ...709..191M, 2010ApJ...709...27W}.  
To reflect these observations in our simulation, only 20\% of the molecular cloud  particle mass, $M_{\rm MC}$, turns into an actual stellar mass, $M_{\rm star}(t)$, over 12 $t_{\rm dyn}$ by
\begin{eqnarray}
M_{\rm star}(t) &=& 0.2 \, M_{\rm MC} \int_0^{\tau} \tau\,' e^{-\tau\,'}\,d\tau\,'  \\
&=& 0.2 \, M_{\rm MC} \left[ 1- (1+ \tau) e^{-\tau} \right],
\label{eq:SF}
\end{eqnarray}
where $\tau = (t-t_{\rm cr})/t_{\rm dyn}$.  
In this formulation, the production of the stellar mass peaks at $t_{\rm dyn}$.  
As $7.5 \times 10^{-7}$ of the rest mass energy of $M_{\rm star}$ is gradually deposited into the cell in which the particle resides,\footnotemark\, this thermal stellar feedback replenishes the energy loss to radiative cooling. 
At the same time, the rest of the molecular cloud particle mass, 0.8\,$M_{\rm MC}$, slowly returns to the gas grid.  
This again reflects the fact that most of the gas in GMCs does not end up locked in stars in a few dynamical time, but is blown out into the ISM to be recycled.  
Meanwhile, 2\% of the ejected mass is counted as metals, contributing to the metal enrichment of the ISM (See Figure \ref{fig:SF}).    

Overall, our feedback treatment corresponds to the energy of $10^{51}$ ergs for every 750 \,\,$M_{\odot}$ of actual stellar mass formed.
Although Type II supernovae explosions are its dominant source \citep{1990ARA&A..28...71S, 2006ApJ...641..878T, 2008ApJ...673..810T}, this feedback also models various other types such as protostellar outflows \citep{2006ApJ...640L.187L, 2008ApJ...687..354N}, photoionization \citep{1989ApJ...345..782M}, and stellar winds \citep{2001dge..conf..181O}.  
Therefore no explicit time delay is necessary between the formation of a molecular cloud and the start of stellar feedback.  
This thermal feedback heats the mass of $\sim10^4 M_{\odot}$ in a $<$ 30 pc cell up to $\sim10^7 \,\,{\rm K}$, but a multiphase medium \citep{1977ApJ...218..148M} is naturally established without using any sub-resolution model.  
The so-called overcooling problem \citep{1999MNRAS.310.1087S, 2001MNRAS.326.1228B} is absent in our simulation since the cooling time of these hot cells is much longer than the sound crossing time \citep{2009ApJ...694L.123K}. 
\footnotetext{Assuming the Salpeter initial mass function $d{\rm n}/dM \propto (M/M_{\odot})^{-2.3}$ in a star cluster \citep{1955ApJ...121..161S}, the fractional mass which ends as Type II supernova (SNII, $> 9\,\,M_{\odot}$) is 1.2\%.  Thus, fixing the mass of each SNII to be $9\,\,M_{\odot}$ we inject $10^{51}$ergs per $9\,\,M_{\odot}$/1.2\% = $750\,\,M_{\odot}$ of the stellar mass formed.  This ratio $10^{51}$ergs\,/\,$750 \,\,M_{\odot} = 1.3\times 10^{48} \,\,{\rm ergs}\,\, M_{\odot}^{-1}$ equals to $7.5 \times 10^{-7}$ of the stellar rest mass energy.}

\begin{figure*}[t]
\epsscale{1.09}
\plotone{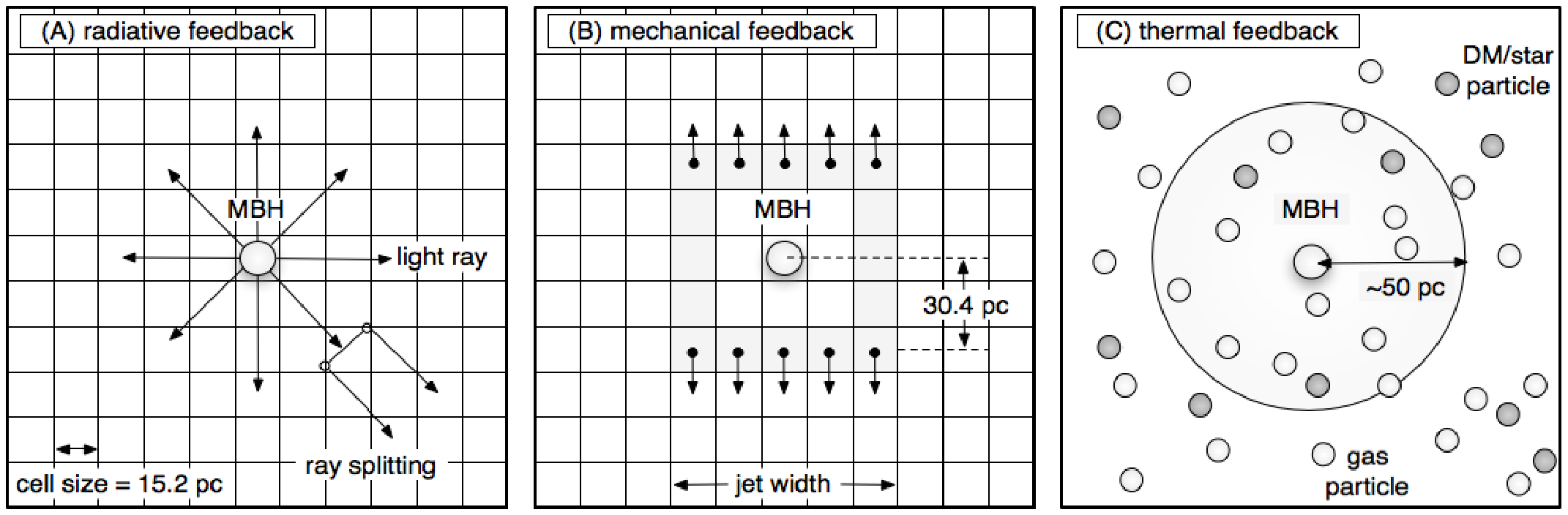}
\caption{Two dimensional schematic views of the different modes of massive black hole feedback. (A) radiative feedback model described in \S \ref{sec:2-RF}: photon rays carrying the energy are adaptively traced via full radiative transfer, (B) mechanical feedback model described in \S \ref{sec:2-MF}: a momentum is injected to the cells around the MBH along pre-calculated directions, and (C) thermal feedback model predominantly used in particle-based galactic scale simulations: thermal energy is  kernel-weighted to the neighboring gas particles around the MBH.
\label{fig:RMTF}}
\end{figure*}

\subsection{Accreting Massive Black Hole} \label{sec:2-accretion}

A $10^5 M_{\odot}$ massive black hole (MBH) is put as a seed at the center of each simulated galaxy.  
It is treated as a collisionless sink particle, but grows in mass by accreting gas from its surroundings.  
We estimate the rate of accretion by employing the Eddington-limited spherical Bondi-Hoyle formula \citep{1944MNRAS.104..273B, 1952MNRAS.112..195B}:
\begin{eqnarray}
\dot M_{\rm BH} &=& {\tt min} (\dot M_{\rm B} \,,\, \dot M_{\rm Edd}) \\
&=& {\tt min} \left( {4 \pi G^2 M^2_{\rm BH} \rho_{\rm B}  \over  { c^3_{\rm s} }    } \,\,,\,\,  {4 \pi G M_{\rm BH} m_{\rm p}  \over  { \epsilon_{\rm r} \sigma_{\rm T} c } }  \right), 
\end{eqnarray}
where $M_{\rm BH}$ is the mass of a MBH, $c_{\rm s}$ is the sound speed of the gas at the cell the MBH resides in, $m_{\rm p}$ is the mass of a proton, and $\sigma_{\rm T}$ is the Thomson scattering cross-section.  
Note that, when compared with Eq.(\ref{eq:Springel}) the nondimensional parameter $\alpha$ is absent.
$\rho_{\rm B}$ is the density at the Bondi radius
\begin{eqnarray}
R_{\,\rm B} = { 2 G M_{\rm BH} \over c^2_{\rm s}} \,\, \simeq \,\, 8.6 \,\,{\rm pc} \left({ M_{\rm BH} \over 10^5 M_{\odot}} \right) \left( { 10 \,{\rm km/s}  \over  c_{\rm s}  }\right)^{2},
\label{eq:Bondi}
\end{eqnarray}
and is extrapolated from the density $\rho_{\rm gas}$ of the cell of size $\Delta x$ where the MBH resides by
\begin{eqnarray}
\rho_{\rm B} = \rho_{\rm gas} \,\cdot\, {\tt min} ((\Delta x/R_{\rm B})^{1.5}, 1.0)  < \rho_{\rm gas}.
\end{eqnarray}
Here an $r^{-3/2}$ density profile is assumed inside the sphere of $R_{\rm B}$ \citep{2010ApJ...709...27W}.
Adopting a radiative efficiency $\epsilon_{\rm r} = 0.1$ for a non-rotating Schwarzschild black hole \citep{1973A&A....24..337S, 2009MNRAS.398...53B}, the Eddington rate for a $10^5 M_{\odot}$ black hole is $\simeq 0.002 \,\,M_{\odot}\, {\rm yr^{-1}}$.  
For more discussion on the accretion rate adopted here, see Appendix \ref{sec:appendix-A}.
To minimize any numerical artifacts, the gas mass accreting onto the MBH is uniformly subtracted from grid cells within a Bondi radius.  
The MBH  also inherits the momentum of the accreting gas.  

Most importantly, to probe the gas dynamics accreting onto the MBH and to fully incorporate the MBH in a galactic simulation, it is imperative to always reach the resolution close to the Bondi radius around the MBH.  
To resolve the gas around the MBH with the best resolution available, eight nearby cells close to the MBH are required to successively refine down to 15.2 pc (proper) at all times.  
In practice, the MBH naturally sits at the densest region most of the time, surrounded by many finest cells.    
While our spatial resolution is still slightly too large to resolve the Bondi radius of a $10^5 M_{\odot}$ black hole, Eq.(\ref{eq:Bondi}), it is enough to resolve the Bondi radii of more massive MBHs such as in nearby X-ray luminous galaxies \citep[e.g. $\sim120\,{\rm pc}$ for SMBH in M87;][]{2006MNRAS.372...21A}.
This shows that our simulations are beginning to depict the self-consistent coevolution of both galaxies and MBHs in one comprehensive framework.  
Admittedly, this resolution is still far from the Schwarzschild radius of any black hole
\begin{eqnarray}
R_{\,\rm Sch} = { 2 G M_{\rm BH} \over c^2} \,\, \simeq \,\, 10^{-8} \,{\rm pc} \left({ M_{\rm BH} \over 10^5 M_{\odot}} \right),
\end{eqnarray}
which is needed to thoroughly describe its accretion disk.
Due to our resolution limit, a MBH particle in our framework represents not just the black hole itself, but also includes accreting gas and stars deep within the galactic nucleus; in other words, the Bondi-Hoyle accretion estimate does not accurately model the physics below the resolution limit (See \S \ref{sec:5-future}).  

\subsection{MBH Radiative Feedback}\label{sec:2-RF}

We now turn our attention to the feedback of an accreting massive black hole.  The gravitational potential energy of the gas accreting onto a black hole is extracted during the gravitational infall.  Assuming an infall down to the innermost stable orbit of an accretion disk, the conversion rate from the rest mass energy to feedback energy is 10\%, previously defined as the radiative efficiency $\epsilon_{\rm r}$.  Hence the bolometric radiation luminosity of a MBH is
\begin{eqnarray}
L_{\rm BH} = \epsilon_{\rm r} \dot M_{\rm BH} c^2 .
\label{eq:luminosity}
\end{eqnarray}
As was discussed earlier, for a long time a {\it thermal} energy deposition has been the dominant strategy to treat the feedback of an accreting MBH \citep[][See \cite{2008MNRAS.387.1403S} or \cite{2011MNRAS.412.1341D} for other approaches]{2005MNRAS.361..776S, 2005Natur.433..604D, 2008MNRAS.387.1163C, 2009MNRAS.398...53B,2010arXiv1002.1712C, 2010arXiv1003.4744T}.
Without question, it has been an effective approximation characterizing the impact of an accreting MBH on a resolved scale when sufficient resolution or full radiative transfer is inaccessible \citep[See Figure \ref{fig:RMTF}(C);][]{2005MNRAS.361..776S}.  
Despite its practical efficiency, however, better feedback models are imperative for high-resolution galaxy formation studies where the Bondi radius is starting to be resolved.  
In the next two sections, we explain the detailed  implementations of two modes of MBH feedback: radiative and mechanical.  
The thermal feedback model previously used can be regarded as an approximation of these two feedback channels combined.  

\vspace{2 mm}

Although the radiation from the MBH in a galaxy was tested in spherically symmetric or axisymmetric models \citep{2007ApJ...665.1038C, 2008ApJ...676..101P, 2009ApJ...699...89C, 2009ApJ...707..823K, 2010arXiv1006.1302P}, a three dimensional radiative transfer calculation of the impact of a MBH has never been performed in  galactic scale simulations.  
In what follows, we treat the MBH as a point source of radiation and carry out a three dimensional transport computation to evolve the radiation fields (See Figure \ref{fig:RMTF}(A); note that molecular cloud particles are not treated as radiation sources).
Achieving high resolution around the MBH is critical here because, if otherwise, the optical depth of the radiation could be small even at the smallest resolved distance from the MBH \citep{2004MNRAS.348.1105O}.  

{\it Enzo}'s radiative transfer module incorporates the adaptive ray tracing technique \citep{2002MNRAS.330L..53A} with the hydrodynamics, energy, and chemistry solvers.  
It has been applied to problems such as the radiative feedback from Pop III stars \citep{2007ApJ...659L..87A, 2008ApJ...685...40W, 2010arXiv1011.2632W} and from Pop III black holes \citep{2009ApJ...701L.133A}.  
For algorithmic and numerical details of {\it Enzo} radiative transfer we refer the readers to \cite{2010arXiv1012.2865W}; and, here we briefly describe the machinery relevant to the presented results.
First the luminosity of the MBH is assigned by Eq.(\ref{eq:luminosity}).
Then 768 ($=12\times4^3$; {\it Healpix} level 3) rays are isotropically cast with a monochromatic energy of $E_{\rm ph} = 2\,\, {\rm keV}$, a characteristic temperature of an averaged quasar spectral energy distribution \citep[SED;][]{2004MNRAS.347..144S, 2005MNRAS.358..168S, 2007ApJ...665.1038C}.\footnote{
A characteristic temperature of the quasar SED is estimated by equating Compton heating and Compton cooling by the given SED \citep{2004MNRAS.347..144S}.  
Therefore it can be considered as the temperature of a Comptonized hot plasma in the vicinity of the MBH, which is represented by our MBH particle resolved only by 15.2 pc. 
Note, however, that the choice of the monochromatic photon energy, $E_{\rm ph}$,  does not change the total luminosity of the MBH, nor does it greatly affect the nonrelativistic Klein-Nishina cross section for Compton scattering, $\sigma_{\rm KN}$, in the regime of $E_{\rm ph} \ll m_{\rm e} c^2$.
}  
Consequently the number of photons per each initial ray is 
\begin{eqnarray}
P_{\,\,\rm init} = { L_{\rm BH}\, dt_{\rm ph} \over E_{\rm ph} \cdot 768} = {  \epsilon_{\rm r} (\dot M_{\rm BH}dt_{\rm ph}) c^2 \,  \over E_{\rm ph} \cdot 768} 
\end{eqnarray} 
given the photon timestep $dt_{\rm ph}$ which we set as the the light-crossing time of the entire computational domain.  
This choice is justified  because the photons are in a free streaming regime, and the energy deposited by the radiation per timestep is relatively small.  
Each ray is traced at speed $c$ until the ray reaches the edge of the computational domain or most of its photons (99.99995\%) are absorbed.  
It is adaptively split into four child rays whenever the area associated with a ray becomes larger than $0.2\, (\Delta x)^2$ of a local cell.   

\begin{table*}  
\caption{Simulation Suite Description}
\centering     
\begin{tabular}{l l   ||  c  c  c  c  c  c   } 
\hline\hline   
\multicolumn{2}{c ||}{Physics\tablenotemark{a}}  & Sim-SF  & Sim-RF  & Sim-MF & Sim-RMF \\ [1ex] 
\hline      
Molecular cloud formation & (See \S \ref{sec:2-formation}) &\textcircled{}&\textcircled{}&\textcircled{}&\textcircled{}\\    
Stellar feedback & (See \S \ref{sec:2-SF}) &\textcircled{}&\textcircled{}&\textcircled{}&\textcircled{}\\    
Massive black hole accretion  & (See \S \ref{sec:2-accretion}) &\textcircled{}&\textcircled{}&\textcircled{}&\textcircled{}\\ 
Massive black hole radiative feedback  & (See \S \ref{sec:2-RF}) &$\times$&\textcircled{}&$\times$&\textcircled{}\\ 
Massive black hole mechanical feedback & (See \S \ref{sec:2-MF}) &$\times$&$\times$&\textcircled{}&\textcircled{}\\ [1ex] 
\hline
\end{tabular} 
\tablenotetext{1}{\scriptsize For detailed explanation, see the referenced section. $\circ$ = included, $\times$ = not included.}
\label{table:desc}  
\end{table*}

Photons in the emitted ray then  interact with the surrounding gas in three ways: they (1) ionize the gas, (2) heat the gas, and (3) exert momentum onto the gas.  
First, the ray loses its photons when it photoionizes H, He, and ${\rm He}^{+}$ with the respective {\it photoionization} rates of
\begin{eqnarray}
k_{\rm ph, H} &=& { P_{\,\,\rm in} (1-e^{-\tau_{\rm H}}) (E_{\rm ph}Y_{k, \rm H} /E_{i, \rm H}) \over n_{\rm H} (\Delta x)^3 dt_{\rm ph} } \\
k_{\rm ph, He} &=& { P_{\,\,\rm in} (1-e^{-\tau_{\rm He}}) (E_{\rm ph}Y_{k, \rm He} /E_{i, \rm He}) \over n_{\rm He} (\Delta x)^3 dt_{\rm ph} } \\
k_{\rm ph, He^+} &=& { P_{\,\,\rm in} (1-e^{-\tau_{\rm He^+}}) \over n_{\rm He^+} (\Delta x)^3 dt_{\rm ph} }
\end{eqnarray}
where $P_{\,\,\rm in}$ is the number of photons coming into the cell, $\tau_{\rm H} = n_{\rm H} \sigma_{\rm H} dl$ is the optical depth,  $n_{\rm H}$ is the hydrogen number density, $\sigma_{\rm H}$ is the energy-dependent hydrogen photoionization cross-section \citep{1996ApJ...465..487V}, $dl$ is the path length through the cell, and $E_i = 13.6,\,\, 24.6,\,\, 54.4\,{\rm eV}$ are the ionization thresholds for H, He, ${\rm He^+}$, respectively.  
The factors $Y_{k}$ are the energy fractions used for ionization when secondary ionizations are considered \citep{1985ApJ...298..268S}.\footnote{$Y_{k, \rm H} = 0.3908(1-x^{0.4092})^{1.7592}$ and $Y_{k, \rm He}= 0.0554(1-x^{0.4614})^{1.6660}$ are fitted as a function of an ionization fraction $x = n_{\rm H^+}/n_{\rm H,tot} \simeq n_{\rm He^+}/n_{\rm He,tot}$; the effect of secondary ionizations on ${\rm He^+}$ can be ignored.}

Second, the excess energy above the ionization threshold, $E_i$, heats each of the species with the {\it photoheating} rates of
\begin{eqnarray}
\Gamma_{\rm H} = { P_{\,\,\rm in} (1-e^{-\tau_{\rm H}}) E_{\rm ph} Y_{\Gamma} \over n_{\rm H} (\Delta x)^3 dt_{\rm ph} }, \,\,\,\,{\rm etc.}
\end{eqnarray}
where $Y_{\Gamma}$ is the fraction of energy deposited as heat when secondary ionizations are taken into account.\footnote{Note that $Y_{\Gamma}  = 0.9971[1-(1-x^{0.2263})^{1.3163}]$ approaches 0 when the ionization fraction gets close to 0.  In other words, when the ionization fraction is low photons are preferentially used to first ionize the gas rather than to heat the gas.}  
The 2 keV soft X-ray photon can also scatter off and heat an electron resulting in the {\it Compton heating} rate of
\begin{eqnarray}
\Gamma_{\rm C} = { P_{\,\,\rm in} (1-e^{-\tau_{\rm e}}) \Delta E(T_{\rm e}) \over n_{\rm e} (\Delta x)^3 dt_{\rm ph} }
\end{eqnarray}
where $\tau_{\rm e} = n_{\rm e} \sigma_{\rm KN} dl$ is the optical depth, $n_{\rm e}$ is the electron number density, $\sigma_{\rm KN}$ is the nonrelativistic Klein-Nishina cross-section \citep[$\simeq \sigma_{\rm T}$\,;][]{1979rpa..book.....R}, and $\Delta E(T_{\rm e}) = 4 k_{\rm B} T_{\rm e} \cdot (E_{\rm ph}/m_{\rm e} c^2)$ is the nonrelativistically transferred energy to an electron at $T_{\rm e}$ \citep{2001ApJ...551..131C}.  
It should be noted that, in Compton scattering, a photon loses its energy by a factor of $\Delta E(T_{\rm e})/E_{\rm ph}$, but essentially keeps propagating without being absorbed. 
However, in order to model this with {\it monochromatic} photons, we instead subtract $P_{\,\,\rm in} (1-e^{-\tau_{\rm e}}) \Delta E(T_{\rm e})/E_{\rm ph}$ photons from the ray.  This is another way a ray loses its photons while traveling through a cell.  Combined, the total heating rate by absorbed and scattered photons becomes
\begin{eqnarray}
\Upsilon= n_{\rm H} \Gamma_{\rm H} + n_{\rm He} \Gamma_{\rm He} + n_{\rm He^+}  \Gamma_{\rm He^+} + n_{\rm e} \Gamma_{\rm C}\,.
\end{eqnarray}

Lastly, photons exert outward momentum to the gas when they are taken out from the ray either by photoionization or by Compton scattering.  
It was claimed that the {\it radiation pressure} from the MBH may markedly alter the environment near the MBH, especially within $\sim$ 0.1 kpc in radius \citep{1995MNRAS.273..249H, 2011MNRAS.412.1341D}.  The large-scale galactic wind driven by deposited photon momentum is also considered as a possible explanation for the $M_{\rm BH}-\sigma_{\rm bulge}$ relation \citep{2005ApJ...618..569M}.  
The added acceleration onto the cell by the radiation pressure is calculated by
\begin{eqnarray}
{\bold a}_{\rm ph} = {d{\bold p}_{\rm ph} \over m_{\rm cell} dt_{\rm ph}} = { P_{\,\,\rm lost} E_{\rm ph} \over \rho_{\rm gas} (\Delta x)^3 c dt_{\rm ph} } {\hat {\bold r}}
\end{eqnarray}
where $d{\bold p}_{\rm ph}$ is the photon momentum exerted onto the cell in $dt_{\rm ph}$, $P_{\,\,\rm lost}$ is the number of photons lost in the cell, and ${\hat {\bold r}}$ is the directional unit vector of the ray.
Neglecting the radiation pressure on dust grains is conservative because its inclusion would further enhance the negative feedback effect (See \S \ref{sec:5-future}).  

\begin{figure*}[t]
\epsscale{1.08}
\plotone{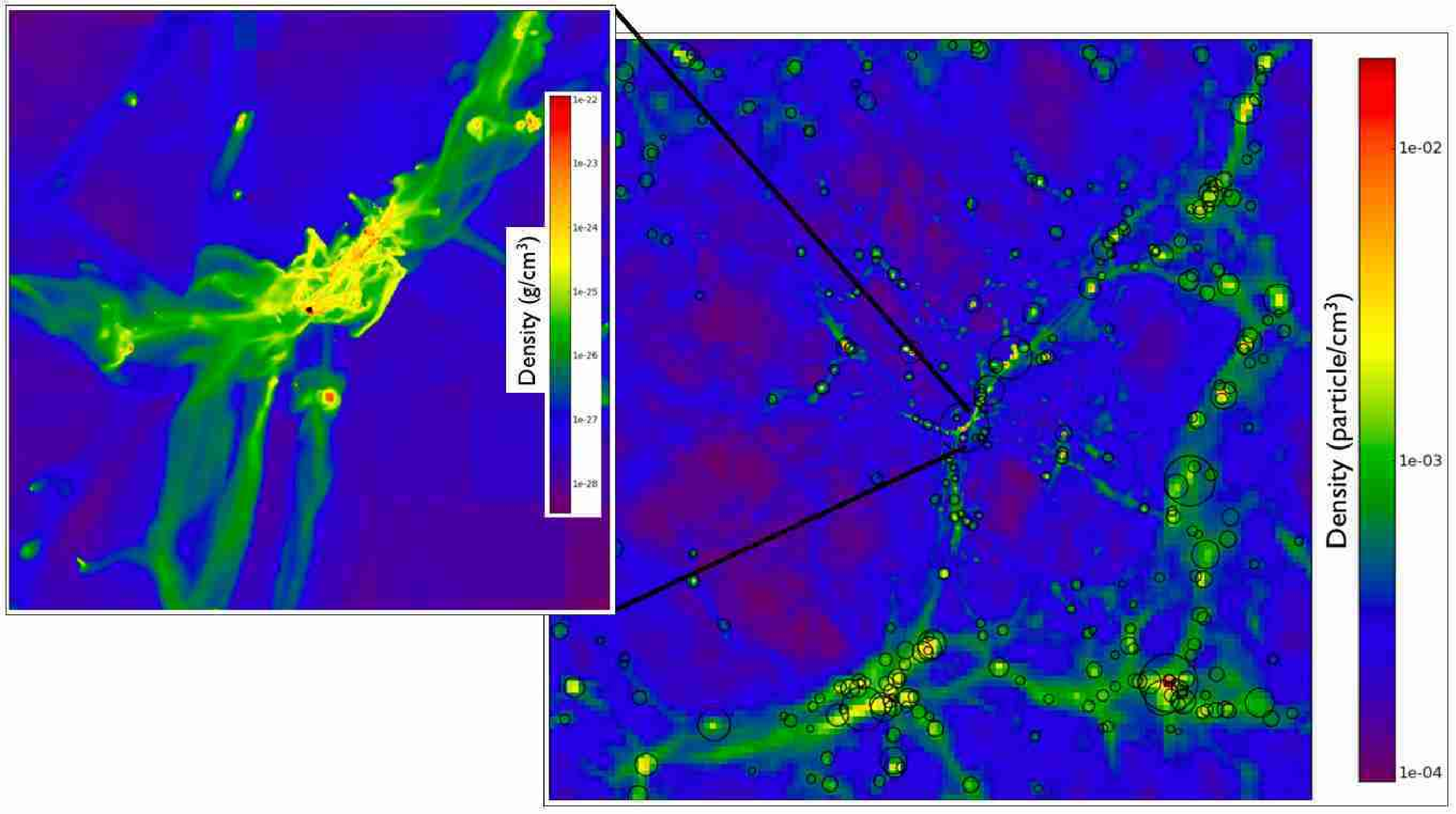}
\caption{A projected density of the simulation box (16 comoving Mpc) at $z=3$ is displayed on the right; circles represent the  identified massive halos.  On the left a $9.2\times10^{11} M_{\odot}$ halo, i.e. the {\it model galaxy}, is shown in a 200 kpc box (proper).  High-resolution images are at http://www.jihoonkim.org/.
\label{fig:initialcondition}}
\end{figure*}

\subsection{MBH Mechanical Feedback}\label{sec:2-MF}

Observations find that a significant portion of the energy extracted during the accretion onto a MBH is released as mechanical energy, creating bipolar jets \citep{1984ARA&A..22..319B, 2003MNRAS.345..705P} or inflating cavities \citep{2002MNRAS.331..369F, 2005Natur.433...45M} at the sites of active galactic nuclei (AGN).   
A number of authors have used a numerical approach to explore the effectiveness of jets in heating up a cooling flow \citep{1994ApJ...436L..63F, 2006PhR...427....1P}; most of them targeted the gas dynamics in galaxy clusters with $\sim$ kpc resolution excluding detailed galactic scale physics \citep[e.g.][]{2004MNRAS.348.1105O, 2007MNRAS.376.1547C, 2008MNRAS.389.1750A, 2010arXiv1004.1851D}.  
In the meantime, a numerical analysis on stellar winds from nuclear disk or MBH jets has been carried out in a galactic scale, but only in an one dimensional context \citep{2009ApJ...699...89C, 2010ApJ...717..708C, 2010ApJ...711..268S} 
Here, we construct a mechanical feedback model of a MBH applicable in three dimensional galactic simulations, which creates accretion-rate-dependent subrelativistic bipolar jets launched at the vicinity of the MBH (See Figure \ref{fig:RMTF}(B)). 

Let us assume that all of the bolometric luminosity of the MBH, $L_{\rm BH}$, is converted to the ``mechanical'' power of jets. 
Because the ejecta has to climb out of the potential well of the MBH, the ``kinetic'' power of the jets is less than $L_{\rm BH}$ by
\begin{eqnarray}
L_{\rm BH}  &=& P_{\rm mech}   \\
&=&  P_{\rm kin} + {\rm (gravitational \,\, potential \,\, energy)}.  
\end{eqnarray}
Therefore the ``kinetic'' power of the jets, as we introduce at a scale of $R_{\rm jet} = 2 \Delta x = 30.4\,\, {\rm pc}$,  can be written as 
\begin{eqnarray}
P_{\rm kin} = \epsilon_{\rm kin} L_{\rm BH} = \epsilon_{\rm kin} \epsilon_{\rm r} \dot M_{\rm BH} c^2 = {1\over 2} \dot M_{\rm jet} v_{\rm jet}^2 ,   \label{eq:jet}
\end{eqnarray}
where $ \epsilon_{\rm kin} < 1$ is the ``kinetic'' coupling constant denoting the fractional energy available for the kinetic motion of the jets (See Figure \ref{fig:RMTF}(B)).  
$\dot M_{\rm jet}$ is the mass ejection rate of the jets, and $v_{\rm jet}$ is the jets velocity when introduced in the simulation. 
Hence $\epsilon_{\rm kin}$ encapsulates not only the acceleration of the jets powered by the AGN central engine, but also the gravitational ``redshift'' from the scale of an accretion disk ($\sim R_{\rm Sch}$) to a resolved scale of jets in simulations ($\sim R_{\rm jet}$).  
\cite{2009ApJ...699...89C} provides estimates for a MBH of $l = \dot M_{\rm BH} / \dot M_{\rm Edd} = 0.005$ as
\begin{eqnarray}
\epsilon_{\rm kin} &=& {P_{\rm kin} \over \epsilon_{\rm r} \dot M_{\rm BH} c^2} = {0.0125 \over \epsilon_{\rm r}(1+400l)^4} \simeq 0.0015\\
\eta_{\rm jet} &\equiv& {\dot M_{\rm jet} \over \dot M_{\rm BH}} = {0.2 \over (1+100l)^4} \simeq 0.04,
\end{eqnarray} 
based on which we fiducially adopt conservative values of $\epsilon_{\rm kin} = 10^{-4}$ and $\eta_{\rm jet} = 0.05$.  

With $\epsilon_{\rm kin}$ and $\eta_{\rm jet}$ now fixed, the kinetic motion of the jets can be fully described. 
First, as usual, out of a sphere of $R_{\rm B}$ centered on the MBH the accreting mass is taken out at every finest hydrodynamical timestep $dt$;  then 5\% of the accreted mass, $\dot M_{\rm jet}dt = 0.05\,\,\dot M_{\rm BH} dt$, is set aside as a mass of jets.  
Now Eq.(\ref{eq:jet}) yields the initial jet momentum, $(\dot M_{\rm jet}dt) v_{\rm jet}$, with
\begin{eqnarray}
v_{\rm jet} = c \left( { 2\epsilon_{\rm kin} \epsilon_{\rm r} \over \eta_{\rm jet} } \right)^{1/2}  = 6000\,\, {\rm km\, s^{-1}}
\end{eqnarray}
for $\epsilon_{\rm r} = 0.1$.
This value of $v_{\rm jet}$ is consistent with numerous observational evidence \citep[e.g.][]{1995PNAS...9211364B, 1999Natur.401..891J, 2009ApJ...706.1253H} and relativistic MHD simulations \citep[e.g.][]{2003ApJ...596.1080V, 2004ApJ...605..656V} suggesting the existence of at least mildly relativistic AGN jets on scales of 1 - 10 pc from a central engine.  
This $v_{\rm jet}$ is also well-matched with the velocity of momentum-driven AGN winds discussed by \cite{2009MNRAS.tmp.1914K}.
Finally the launch speed of the surrounding cells is found by averaging the momentum of jets and the preexisting gas in those cells \citep{2010ApJ...709...27W}.  

One may want to continuously launch the jets at every finest hydrodynamical timestep.  
However, if the injected mass of jets is minuscule compared to the preexisting mass in surrounding cells, the jets make little or no dynamical impact on the surrounding cells after being mass-weighted averaged with them.  
Since it is unfeasible to resolve all gas cells around the MBH down to $\dot M_{\rm jet} dt$, an alternative approach is indispensable. 
Moreover, there is growing observational evidence of double-lobed radio galaxies (or double-double radio galaxies; DDRG) implying that the jets have launched in an episodic fashion with jets interruption timescales of $10^5$ -$10^8$ years \citep{2004ApJ...613..119S, 2006MNRAS.366.1391S}.  
These two considerations lead us to adopt the following method: every time the accumulated jet mass, $\Sigma \,\dot M_{\rm jet} dt$, exceeds the threshold of $300\,\, M_{\odot}$ it is injected in collimated bipolar jets of a width of five finest cells in the vicinity of the MBH.  
This approach renders jets intermittent (once every 30 Myr if $\dot M_{\rm BH} = 10^{-5} M_{\odot} {\rm yr}^{-1}$) and dynamically important in our calculation.  

The jets are injected parallel and anti-parallel to the total angular momentum $\bold L$ of the accreted gas up to that point.  
The angular momentum vector $\bold L$ changes its direction frequently while it asymptotes to the overall galactic rotation axis.  
This implementation is motivated by the observations of X-shaped radio galaxies (XRGs) where the radio jets rapidly reorient themselves by the interaction with the surrounding gas or by mergers \citep{2002Sci...297.1310M, 2003ApJ...594L.103G}.  
Lastly,  since the mechanical or the radiative feedback alone may not describe the whole picture, we include hybrid models in which each of these two channels constitutes half of the MBH bolometric luminosity, $L_{\rm BH}$ (Sim-RMF; see Table \ref{table:desc}).  

Note that the mechanical channel has not been a main driver of MBH feedback in the presented calculation because, with highly suppressed mass accretion rate, jets have launched only a few tens of times in 350 Myr (See \S \ref{sec:4-gas}).  We later comment upon its efficiency in \S \ref{sec:5-future}.

\section{Initial Conditions}\label{sec:3}

These improved physics of galaxy formation are first extensively tested in isolated galaxies.  
We then apply them to a massive star-forming galactic halo of $9.2\times10^{11} M_{\odot}$ at redshift 3 in a cosmological $\Lambda$CDM simulation.  We begin by describing how the initial conditions of our simulation are generated.

\subsection{Setting up a $\sim{\it 10^{12}} M_{\odot}$ halo} \label{sec:3-setup}

A three dimensional cubic volume of 16 comoving Mpc on a side is set up at $z=60$ assuming a flat $\Lambda$CDM cosmology with dark energy density $\Omega_{\Lambda} = 0.73$, matter density $\Omega_{\rm m} = 0.27$, baryon density $\Omega_{\rm b} = 0.044$, and Hubble constant  $h=0.71$ (in the unit of $H_0 = 100 {\, \rm km\,\, s}^{-1} {\rm Mpc}^{-1}$).  
A scale-invariant primordial power spectrum \citep[spectral index $n = 1$,][]{1999ApJ...511....5E} is adopted with  $\sigma_8 = 0.81$, the rms density fluctuation amplitude in the sphere of 8 $h^{-1} {\rm Mpc}$.  

We identify a dark matter halo of $\sim10^{12} M_{\odot}$ at $z=3$ by performing a coarse-resolution adiabatic run.  
Then we recenter the density field around this halo and set up a new initial condition which preserves the same large-scale power yet contains a small-scale power as well, with a $128^3$ root grid and a series of two nested child grids of twice finer resolution each ($160^3$ cells for level $l=1$, and $200^3$  for $l=2$).  
Therefore the finest nested grid at level $l=2$ spans 6.25 comoving Mpc on a side, contains $200^3$ dark matter particles of ${9.6 \times10^{5} M_{\odot}}$, and manifests the equivalent resolution of a $512^3$ unigrid.  
Initially all the cells throughout $l=2$ grids are allowed to be further refined;  however, the volume in which additional refinement is enabled (${\bold V}_{\rm ref}$\,; in the shape of a rectangular solid) continually shrinks in size in such a way that it encloses only the smallest dark matter particles.\footnotemark\, 
An initial metallicity of $Z=0.003\,Z_{\odot}$ is also set up everywhere to track the metallicity evolution and to facilitate cooling below $10^4$ K. 
\footnotetext{This active adjustment on the size of ${\bold V}_{\rm ref}$ prevents heavier dark matter particles of initial $l=0$ and $l=1$ grids from penetrating the central region of a simulation box, thereby causing runaway refinement.  
Typically ${\bold V}_{\rm ref}$ becomes $\sim 60\%$  of the entire $l=2$ region in length at $z=3$, which is still large enough to encompass the Lagrangian volume of a $\sim 10^{12} M_{\odot}$ halo at $z=3$.}

Our initial condition is first evolved to $z = 3$ with a low-resolution (121.6 pc) refinement strategy and a particle formation and feedback recipe, without an accreting MBH.  
At $z=3$ we split each dark matter and star particle inside the focused volume (${\bold V}_{\rm foc}$; a rectangular hexahedron of 1.28 comoving Mpc on a side, a subset of ${\bold V}_{\rm ref}$)  into 13 child particles using the particle refinement technique by \cite{2002MNRAS.330..129K}.  This algorithm places child particles on a hexagonal close packed (HCP) array, and has been applied to many particle-based applications requiring enhanced particle resolution in a resimulated region \citep[e.g.][]{2003ApJ...596...34B, 2007MNRAS.378..507K, 2008Sci...321..669Y}.
After the particle splitting procedure, each dark matter particle  in ${\bold V}_{\rm foc}$ represents a collective mass of $74000 \,\,M_{\odot}$.  Across ${\bold V}_{\rm foc}$ cells are now allowed to refine up to 11 additional levels, achieving maximum spatial resolution of 15.2 pc at $z\sim3$ (See \S \ref{sec:2-ref}).    

\subsection{Galactic parameters}

Consequently, this process produces our focused object at $z=3$ dubbed a {\it model galaxy}, on which a suite of high-resolution simulations is performed (Figure \ref{fig:initialcondition}).  
The model galaxy has a mass of $M_{\rm vir} \simeq 9.2 \times10^{11} M_{\odot}$ at $z=3$ and a corresponding virial radius of
\begin{eqnarray} 
R_{\rm vir} = M_{\rm vir}^{1/3} \left[ {H^2_0h^2 \Omega_{\rm m} \Delta_{\rm c} \over 2 G}  \right]^{-{1/3}} \simeq 310  \,\,\,{\rm comoving \,\,\, kpc}
\end{eqnarray}
given $h=0.71$, $\Omega_{\rm m} = 0.27$, and $\Delta_{\rm c} = 200$.  
The dark matter halo represented by $\sim 1.1 \times 10^7$ particles constitutes $\sim 88\%$ of the total mass.  
About $\sim 1.0 \times 10^7$ particles contain ${8.0 \times10^{10} M_{\odot}}$ of stellar mass, whereas the rest, ${3.5 \times10^{10} M_{\odot}}$, is in gaseous form available for future star formation, either in the ISM or in the embedding halo.  
There is no shortage of gas supply, as the gas from outside the halo continuously falls inward either by spherical accretion or by cold accretion along one of the multiple filaments \citep{2009Natur.457..451D, 2010MNRAS.404.2151C}.  
The halo has spin parameters of 0.051 for dark matter, and 0.069 for gas.  
At the center of all lies a $10^5 M_{\odot}$ MBH we plant as a gravitational seed. 
This choice of the initial MBH mass lies below the \cite{1998AJ....115.2285M} relationship assuming 10\% of the stellar mass is in the bulge, which may have resulted in a weaker mode of MBH feedback - possibly a ``radio-mode'' analogue - and the negligible gas expulsion by the MBH (See \S \ref{sec:4-gas} and \S \ref{sec:4-MBH}).  
Therefore the reported results should not be interpreted as a general picture of MBH feedback.  
It remains to be seen whether more massive MBHs or fast growing MBHs have different effects.  
We will come back to this issue  in \S \ref{sec:4-MBH}.  

\section{Results}\label{sec:4}

A suite of simulations with optional modes of feedback is performed from $z=3$ to 2.6 in order to investigate the evolution of a massive star-forming galaxy with its embedded massive black hole.  
We mostly focus on two simulations, one with and the other without MBH feedback (Sim-SF and Sim-RMF; Table \ref{table:desc}).     
Each of the calculations is performed on 16 processors of the Orange cluster\footnote{Infiniband-connected AMD, 8 cores per node, 4 GB memory per core} at Stanford University.   
Grids and particles altogether, each simulation is routinely resolved with $\sim 6.5\times10^7$ total computational elements ($\sim 4.5\times10^7$ particles and $\sim 270^3$ cells).  
To evolve the system for 350 Myr, each of these runs typically takes $\sim 20000$ CPU hours.  

\begin{figure}[t]
\epsscale{1.05}
\plotone{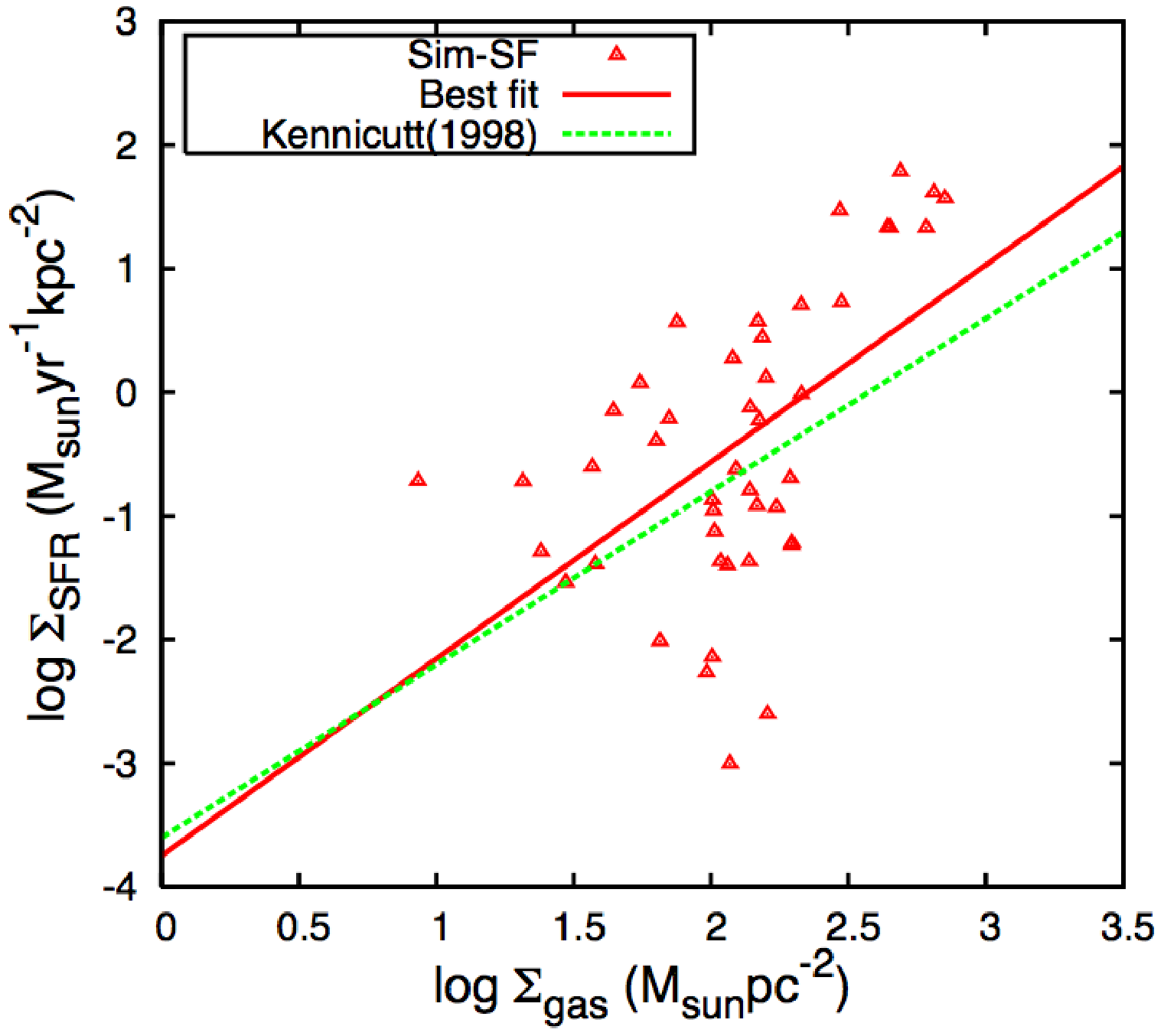}
\caption{The relationship between star formation rate (SFR) and gas surface density.  The data is from a ${\rm (20 \,\,kpc)^3}$ box centered on a MBH in Sim-SF at $z=2.75$.  The solid line is the best fit for simulated data while the dashed line is from observations of nearby galaxies \citep[$\Sigma_{\rm SFR} \propto \Sigma_{\rm gas}^{1.4}$;][]{1998ApJ...498..541K}.
\label{fig:Schmidt}}
\end{figure}

\begin{figure*}[t]
\epsscale{1}
\plotone{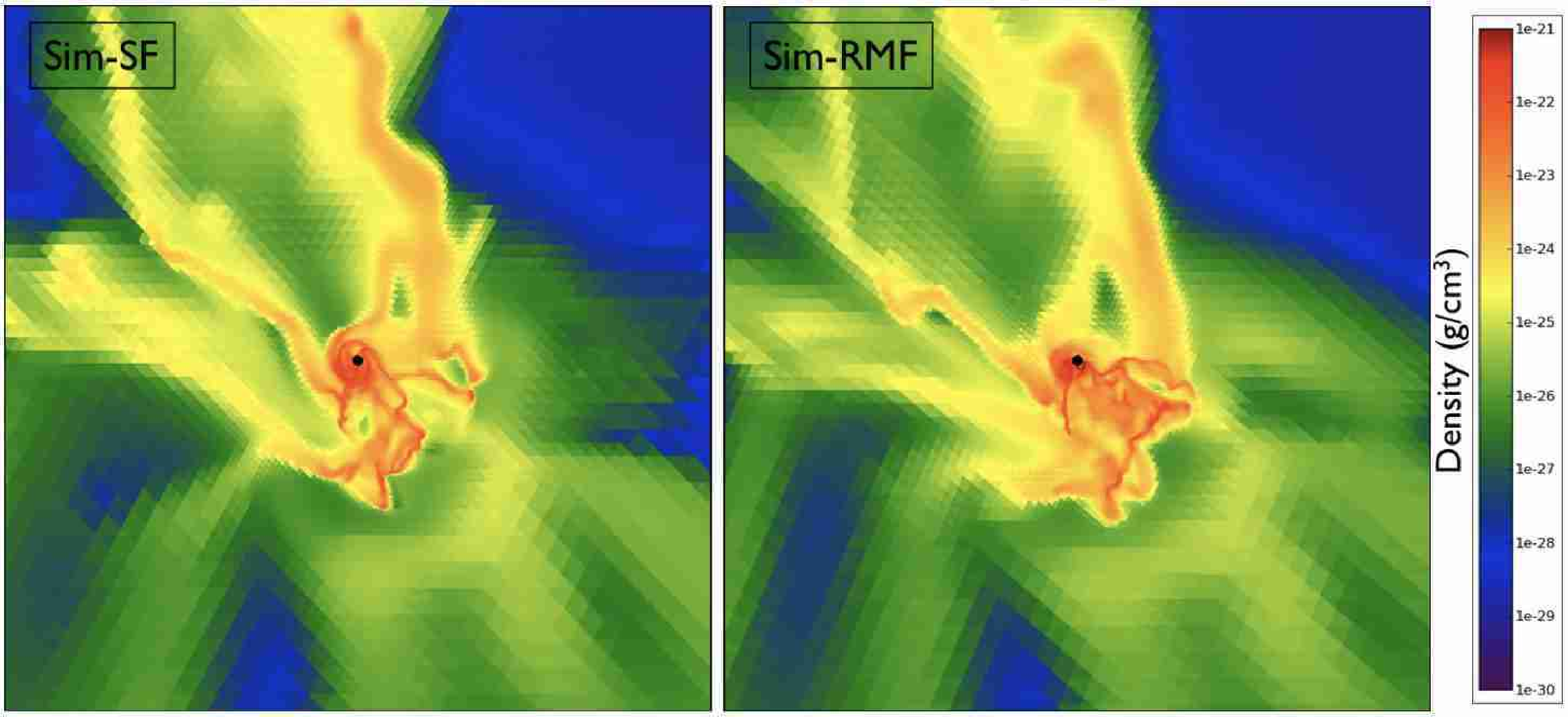}
\caption{The face-on views of the disks.  Density in the central 20 kpc (proper) sliced through the MBH (black dots at the centers) at $z=2.75$, about 220 Myr after the MBH is placed.  Sim-SF on the left, and Sim-RMF on the right.  Compare with Figure \ref{fig:temperature_L}.  
\label{fig:density_L}}
\end{figure*}

\begin{figure*}[t]
\epsscale{1}
\plotone{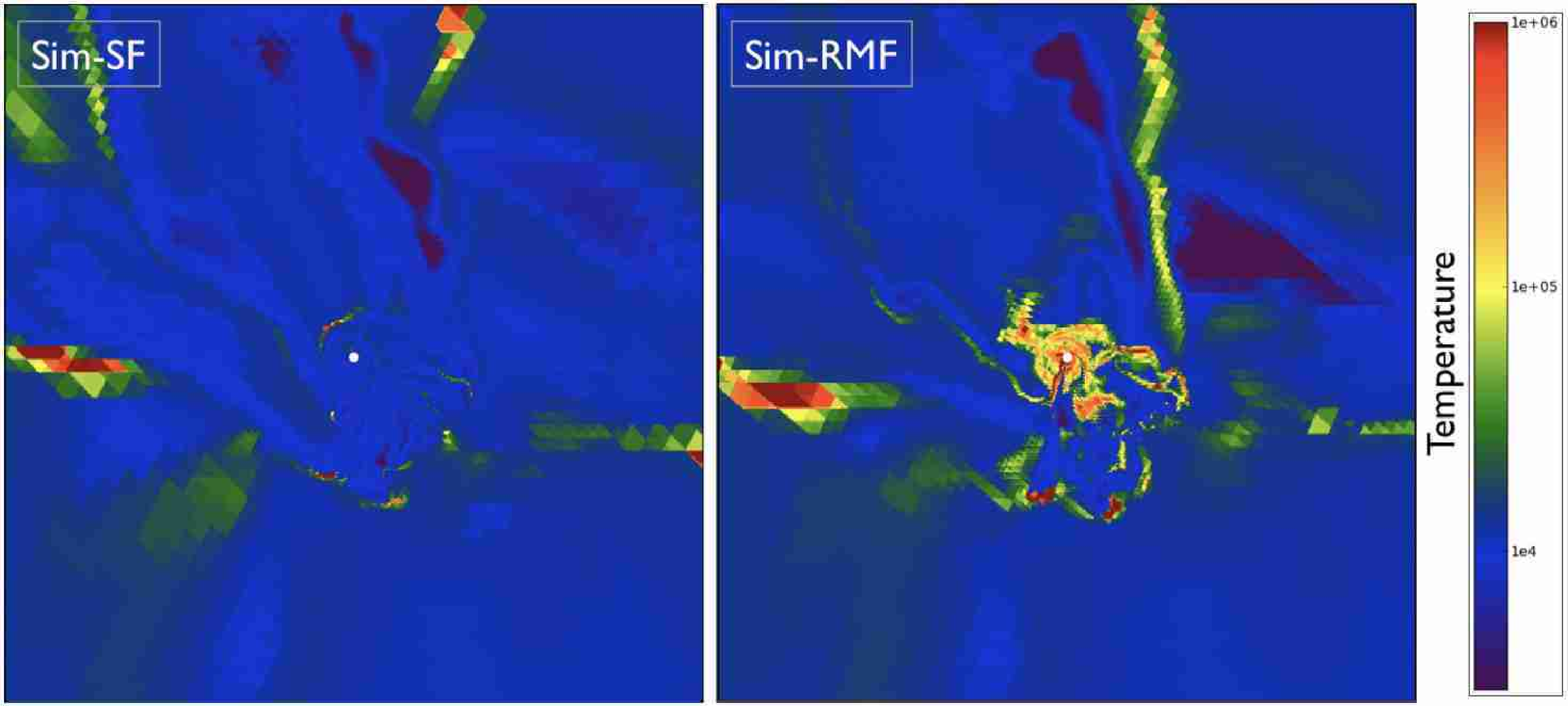}
\caption{The face-on views of the disks.  Temperature in the central 20 kpc (proper) sliced through the MBH (white dots at the centers) at $z=2.75$.  Sim-SF on the left, and Sim-RMF on the right.  A hot region of a size $\sim$ 2 kpc in Sim-RMF heated by MBH feedback is prominent, which remarkably contrasts with a much colder ISM in Sim-SF.
\label{fig:temperature_L}}
\end{figure*}

\begin{figure}[t]
\epsscale{1.12}
\plotone{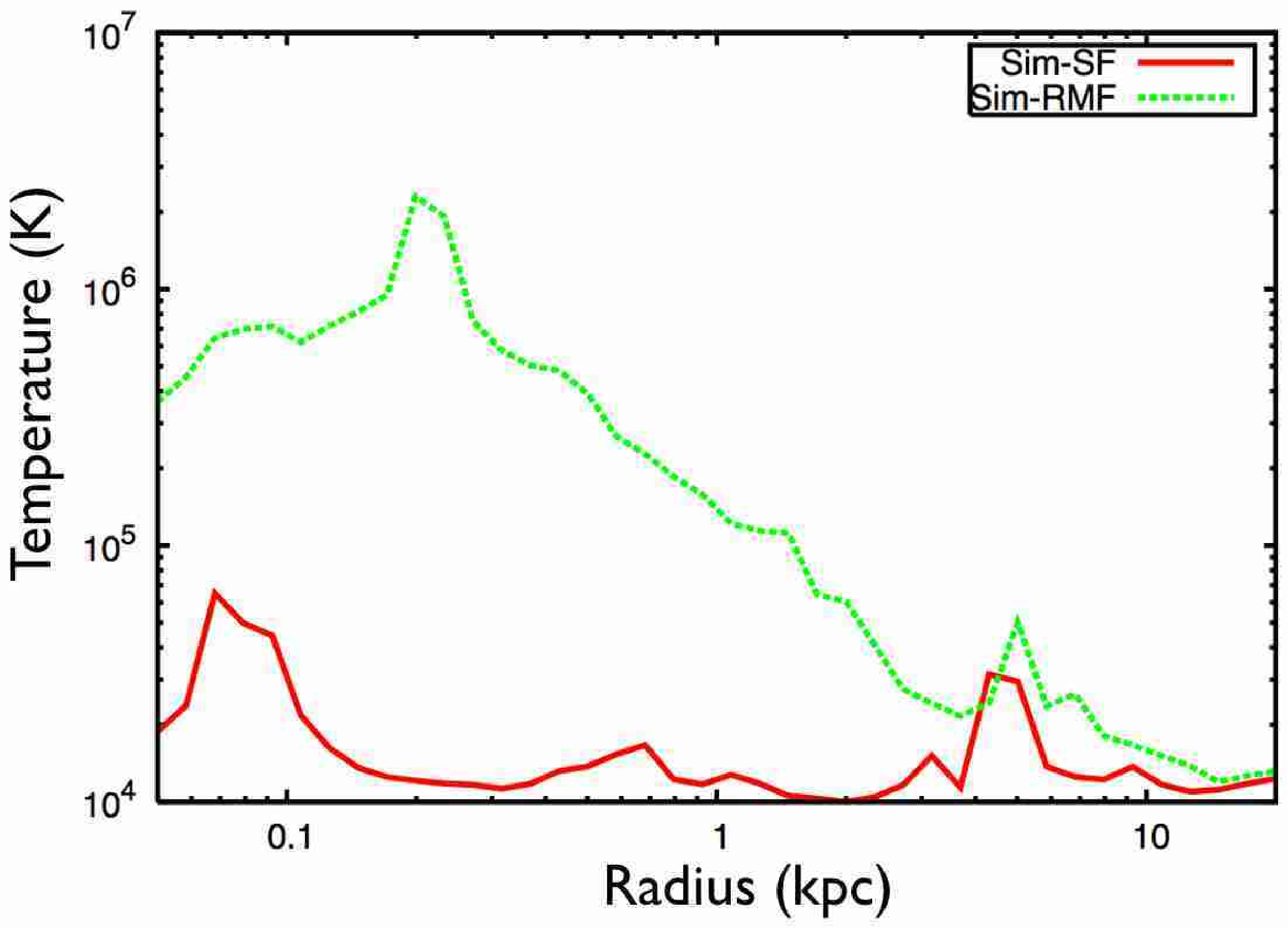}
\caption{The mass-weighted radial profile of temperature in a 20 kpc sphere centered on the MBH at $z=2.75$.  The red solid line and the green dashed line represent Sim-SF and Sim-RMF, respectively.   The temperature within $\sim$ 2 kpc radius is raised mostly by the radiation from the MBH.
\label{fig:prof_temp}}
\end{figure}

\subsection{Star Formation Rates}\label{sec:4-Schmidt}

First we check the validity of our molecular cloud formation criteria (\S \ref{sec:2-formation}) and stellar feedback (\S \ref{sec:2-SF}) by comparing star formation rate (SFR) with gas density.  
Figure \ref{fig:Schmidt} displays a relation between the SFR surface density and the gas surface density in Sim-SF at $z=2.75$.  
In a ${\rm (20 \,\,kpc)^3}$ box centered on a MBH, each data point is made by taking the mean values in a ${\rm (1\,\, kpc)^3}$ bin, the typical aperture size of $\Sigma_{\rm SFR} - \Sigma_{\rm gas}$ studies for spatially-resolved nearby galaxies \citep[e.g.][]{2007ApJ...671..333K}.
Here Eq.(\ref{eq:SF}) is used to calculate the stellar mass newly spawning in each cell, and the data points below the observation limit, $10^{-5} M_{\odot} {\rm yr^{-1} kpc^{-2}}$, are discarded.

The molecular cloud formation and stellar feedback, joined with high spatial resolution, work together to self-regulate star formation.  
However, authors note that our best fit to this particular snapshot of the galaxy is steeper than the observed trend of $z \sim 0$ with larger dispersion.

\begin{figure*}[t]
\epsscale{1.17}
\plotone{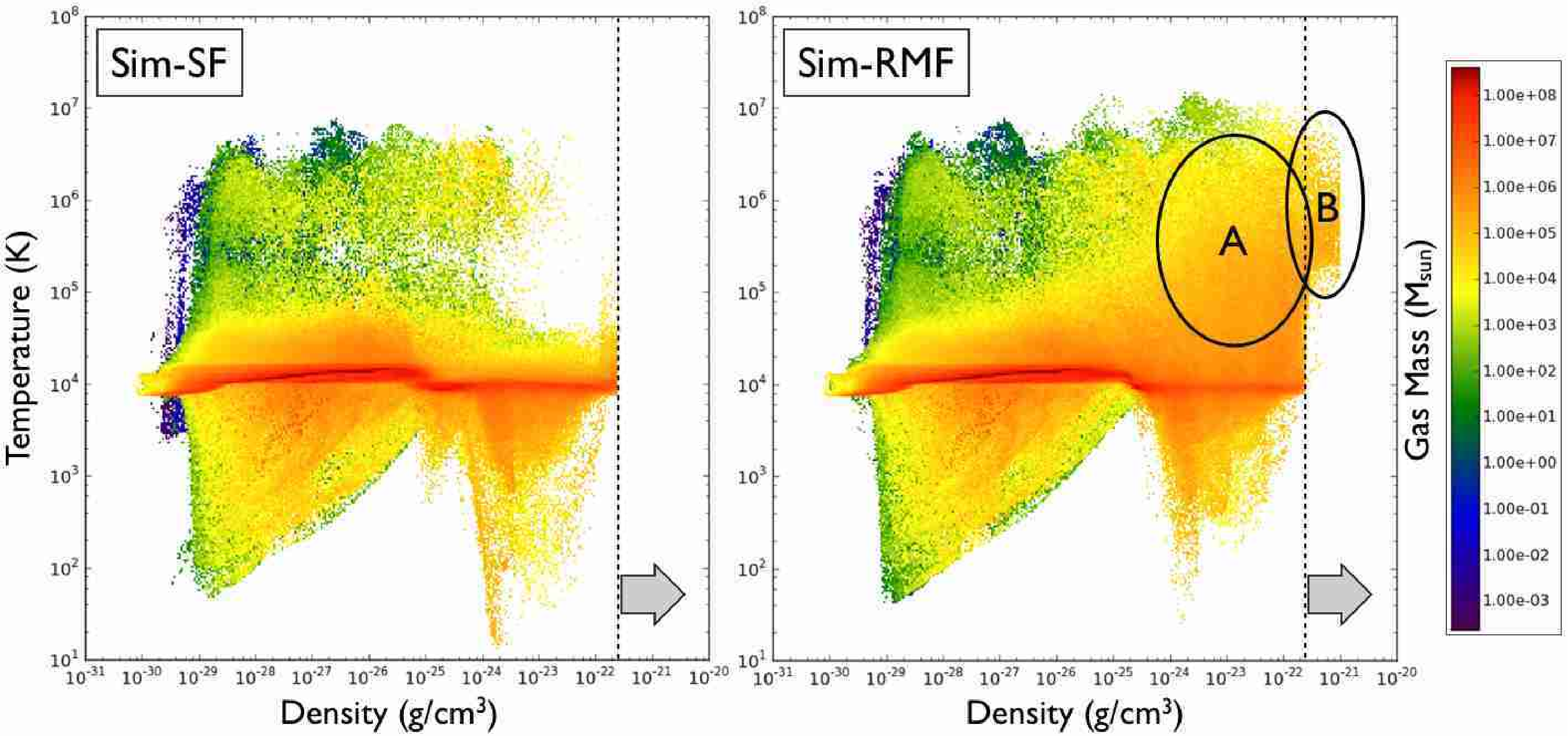}
\caption{Joint probability distribution functions (PDFs) of gas density and temperature colored by gas mass in each bin.  The data is for a 200 kpc sphere centered on the MBH at $z=2.75$.  Sim-SF on the left, and Sim-RMF on the right.  The vertical dashed line in each plot denotes the density threshold for molecular cloud formation $n_{\rm thres}$ (\S \ref{sec:2-formation}).  Note that the MBH feedback in Sim-RMF heats the dense gas up to $10^6$-$10^7$ K increasing the amount of high density gas stable against fragmentation (zone ``A'').. 
\label{fig:PDF}}
\end{figure*}

\subsection{Lack of Star-forming Gas in the Inner Core}\label{sec:4-gas}

Now we turn to the topic of an accreting massive black hole and its feedback.  
We focus on how MBH feedback changes its surrounding ISM, and how it locally suppresses molecular cloud formation.
For this purpose, we hereafter examine the snapshot of the model galaxy at $z=2.75$ (or 2410 Myr after the Big Bang), about 220 Myr after an accreting MBH is placed at the center of the galaxy.  
The model galaxy now has a mass of $M_{\rm vir} \simeq 9.0\times 10^{11} M_{\odot}$ and correspondingly, a virial radius of $R_{\rm vir} \simeq 80\,\,{\rm kpc}$ (proper; radii are hereafter in proper kpc, not comoving, unless marked otherwise). 

When a MBH starts to accrete gas, the gravitational potential energy of the accreting gas is released in the form of radiation and jets.  
Even in the case of a slowly growing MBH, as in our simulations (a possible ``radio-mode'' analogue; $\dot M_{\rm BH} \sim 0.05 \,\dot M_{\rm Edd}$; see \S \ref{sec:4-MBH}), the feedback from the MBH is known to play a major role in regulating star formation and its own growth \citep{2006MNRAS.365...11C, 2007ARA&A..45..117M, 2007MNRAS.380..877S, 2008ApJ...687L..53P}.  

The density and temperature structures in the central regions of the galaxies from Sim-SF (left; without MBH feedback) and Sim-RMF (right; with radiative and mechanical MBH feedback) are shown in Figures \ref{fig:density_L} and \ref{fig:temperature_L}.  
In particular, in Figure \ref{fig:temperature_L} for Sim-RMF, a hot region of size $\sim$ 2 kpc surrounding the MBH is prominent, which remarkably contrasts with a much colder ISM in Sim-SF.  
This region is heated up to $10^6$ K by ionizing photons heating hydrogen and helium, and scattering off electrons. 
The latter, i.e. Compton heating, is important especially in a highly ionized region.  
When the fractions of neutral species (or singly ionized He) are low, the effect of photoheating on $\rm H$, $\rm He$, and $\rm He^+$ is mild; instead, the contribution of Compton heating on electrons is relatively large.  
The hot temperature at the center of Sim-RMF is also evident in the radially averaged temperature profile of Figure \ref{fig:prof_temp}.
Note that the 1-2 kpc distance over which the gas is heated is consistent with the characteristic distance found in an analytic study  \citep[Figure 4 of ][]{2005MNRAS.358..168S} out to which the gas is heated by photoionization and Compton scattering.  

A hot temperature in the inner core of the galaxy in Sim-RMF leads to a significant deprivation of cold, dense star-forming gas.  
Figure \ref{fig:PDF} illustrates how the structure of ISM is changed by MBH feedback, in terms of joint probability distribution functions of gas density and temperature.  
\begin{itemize}
\item The left figure depicts a typical ISM without MBH feedback but still with stellar feedback (Sim-SF).  
It features a multiphase ISM that is naturally achieved in adaptively refined mesh, including cold, dense star-forming gas (T $< 10^4$ K, $\rho >10^{-24} \,\,{\rm g \,\,cm^{-3}}$), and hot diffuse supernovae bubbles.
As expected, above the molecular cloud formation threshold ($n_{\rm thres}=125$ ${\rm cm^{-3}}$; denoted by a dashed line), gas cells immediately turn into molecular cloud particles, and thus no cell is left behind unresolved.  
\item On the right, the gas cells of density $> 10^{-24} \,\,{\rm g \,\,cm^{-3}}$ are now heated up to $10^6$ - $10^7$ K, populating zone ``A".
These cells are mostly located on the disk in relatively close proximity ($<$ 2 kpc) to the central MBH. 
These cells are very stable against fragmentation because they are so hot that they can barely cool down to a typical molecular cloud temperature in a dynamical time (i.e. $t_{\rm cool} \gg t_{\rm dyn}$). 
For that reason, the cells have hard time to fulfill the condition (c) of the molecular cloud formation criteria described in \S \ref{sec:2-formation}.  
The heating by the X-ray radiation thus increases the amount of dense gas in the vicinity of the MBH which is incapable of condensing into stars.\footnote{Some hot gas cells above the molecular cloud formation threshold still exist on this PDF, not turning into particles (zone ``B'').  While it is due to the violation of one of the molecular cloud formation conditions ($t_{\rm cool} < t_{\rm dyn}$), we emphasize that the thermodynamical properties here are unreliable as they are above the resolution-dependent molecular cloud formation threshold.}
\end{itemize}

\begin{figure*}[t]
\epsscale{1.15}
\plotone{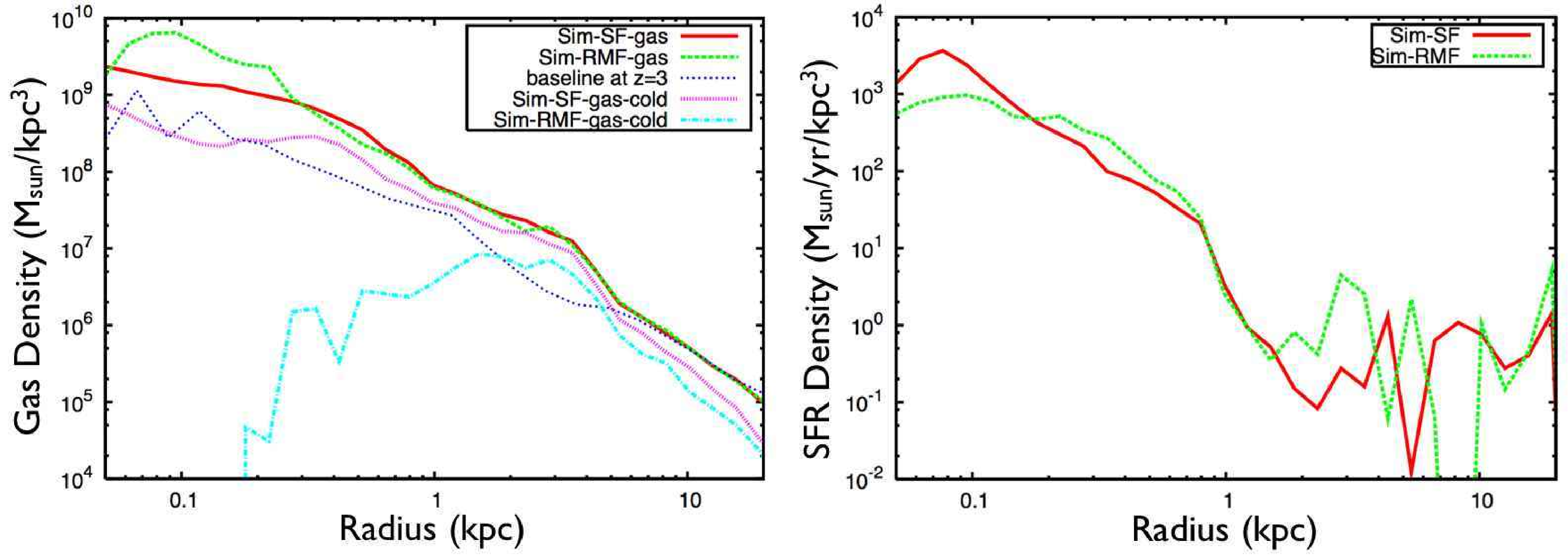}
\caption{Spherically-averaged radial gas density profiles centered on the MBH at $z=2.75$.  In each panel, the red solid line and the green dashed line represent Sim-SF and Sim-RMF, respectively.  Left: in Sim-RMF the radiation from the MBH keeps a large amount of gas at the galactic center (green dashed line) compared to Sim-SF (red solid line).  Only a minimal fraction of the gas within 0.2 kpc radius is below $10^4$ K (the cyan dot-dashed line) whereas in Sim-SF, 10 - 30\% of the gas in the same region is considered to be cold (the pink dotted line).  The blue thin dotted line shows the density profile at $z=3$.  Right: star formation rate (SFR) density shows significantly suppressed star formation activity in the central 0.2 kpc sphere. 
\label{fig:prof_dMdr}}
\end{figure*}

Therefore, the feedback from even a slowly growing MBH retains hot dense gas at a galactic center which otherwise could have created strong star formation.  
Figure \ref{fig:prof_dMdr} dramatically demonstrates the distinct changes in radially averaged density profiles when MBH feedback is included.  
\begin{itemize}
\item The left figure of gas density profiles displays that the radiation from the MBH keeps a substantial amount of gas at the core of the galaxy, and inhibits the gas from all turning into stars.  
The total gas density of Sim-RMF (the green dashed line) at 0.1 kpc from the MBH is about $\sim$ 5 times as high as that of Sim-SF (the red solid line). 
The thin blue dotted line represents the initial profile of gas at z=3 when the simulation restarts with 15 pc resolution.
However in Sim-RMF, only a minimal fraction of the gas within 0.2 kpc radius is below $10^4$ K (the cyan dot-dashed line) whereas in Sim-SF, 10 - 30\% of the gas in the same region is considered to be cold, thus potentially star-forming (the pink dotted line).  
\item The deprivation of cold dense gas in Sim-RMF inevitably prompts the suppression of star formation activity in the inner core of the galaxy.  
The right figure reveals the star formation rate density (in $M_{\odot}{\rm yr^{-1}kpc^{-3}}$) as a function of distance from the MBH.  
Eq.(\ref{eq:SF}) is again used to calculate the new stellar mass being generated in each cell.  
The SFR density of Sim-RMF at 0.1 kpc from the MBH is reduced by more than $\sim$ 50\% when compared with that of Sim-SF.\footnotemark\,\,
Overall, the X-ray radiation from the MBH severely suppresses the SFR of Sim-RMF inside the 0.2 kpc sphere.  
\footnotetext{  
Note that since we used Eq.(\ref{eq:SF}) to estimate SFR, each molecular cloud particle will generate stellar mass for 12 $t_{\rm dyn}$.  
Therefore the SFR inside the 0.1 kpc sphere would include a number of particles that were formed outside the sphere, but have migrated inward and are now forming stars there.  
This is why the SFR density is not as suppressed as one would have naively expected from the density profile of cold gas.}
\end{itemize}

\begin{figure}[t]
\epsscale{1.16}
\plotone{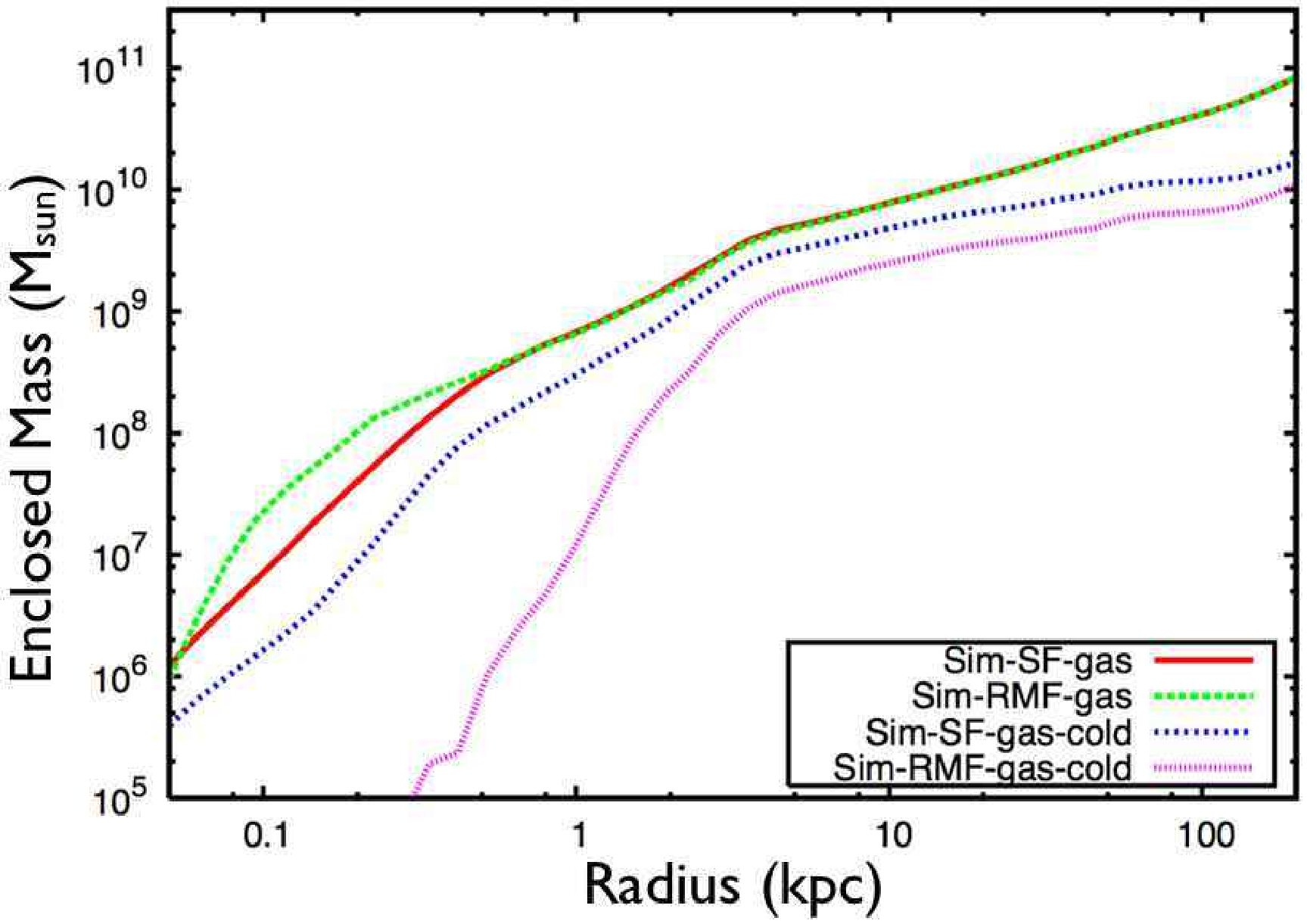}
\caption{Enclosed mass profiles of total gas and cold gas (${\rm T} < 10^4$ K) centered on the MBH at $z=2.75$.  The red solid line and the green dashed line represent Sim-SF and Sim-RMF, respectively.  Sim-RMF harbors more gas within 0.2 kpc than what Sim-SF does, but almost none of this gas is  below $10^4$ K. 
\label{fig:enc_mass_gas}}
\end{figure}

The gas masses enclosed within a radius r, $M_{\rm gas}(<{\rm r})$,  are plotted in Figure \ref{fig:enc_mass_gas}, showing the impact of MBH feedback as a powerful energy source to reshape the galactic gas distribution.
Sim-RMF harbors $\sim$ 2.5 times more gas ($1.3\times10^8 M_{\odot}$) within 0.2 kpc than what Sim-SF does ($5.1\times10^7 M_{\odot}$), but almost none of this gas is below $10^4$ K.  
This gas in the core is not consumed by star formation, nor is pushed away by any mechanical outflow.  
Note that in Sim-RMF cold gas mass within 10 kpc is reduced by $\sim$ 50\% displaying how far the MBH radiation reaches.  

Readers should also note that the enclosed gas masses at the virial radius (80 kpc proper) are almost identical between Sim-SF and Sim-RMF, implying there has been {\it no massive gas expulsion driven by the MBH}.  
This is one of the key differences from previous numerical studies.
Very strong gas expulsions were frequently observed in previous studies in which the MBH feedback energy is only thermally deposited to a few neighboring gas particles \citep[e.g.][]{2005MNRAS.361..776S}.  
In contrast, most of the gas is still bound to our simulated galaxies because (1) the energy released from our slowly growing MBH is relatively small (a possible ``radio-mode'' analogue),\footnote{
This is valid only for the results presented herein in which the mass accretion onto the MBH is highly suppressed by self-regulation.  
It remains yet to be seen whether more massive MBHs or fast growing MBHs (for example, in merging galaxies) do or do not unbind the gas from the galactic gravitational potential.  
The so-called ``quasar-mode'' feedback will be the topic of a future paper (See \S \ref{sec:5-future}).}  
(2) the gas mass to which the MBH energy is coupled is large in our radiative feedback formalism, and (3) the mass accretion rate onto our MBH is not high enough to repeatedly drive jets.
Note again that the mechanical channel of MBH feedback is not a main driver of feedback in the presented calculation.
The MBH has not doubled its mass at the end of our calculation (after 350 Myrs; see \S \ref{sec:4-MBH}); and, with this highly suppressed mass accretion rate, jets have launched only a few tens of times in 350 Myr.  
We come back to the efficiency issue of mechanical feedback in \S \ref{sec:5-future}.

To summarize, we have shown that MBH feedback, especially its radiation, alters the multiphase ISM of the surrounding gas and thus deprives the galactic inner core of cold, dense star-forming gas.  
Two consequences arise from the lack of star-forming gas at the galactic center: locally suppressed star formation, and the associated change in stellar distribution.
We discuss these topics in the following sections.

\subsection{Locally Suppressed Star Formation and the Change in Stellar Distribution}\label{sec:4-star}

\begin{figure}[t]
\epsscale{1.18}
\plotone{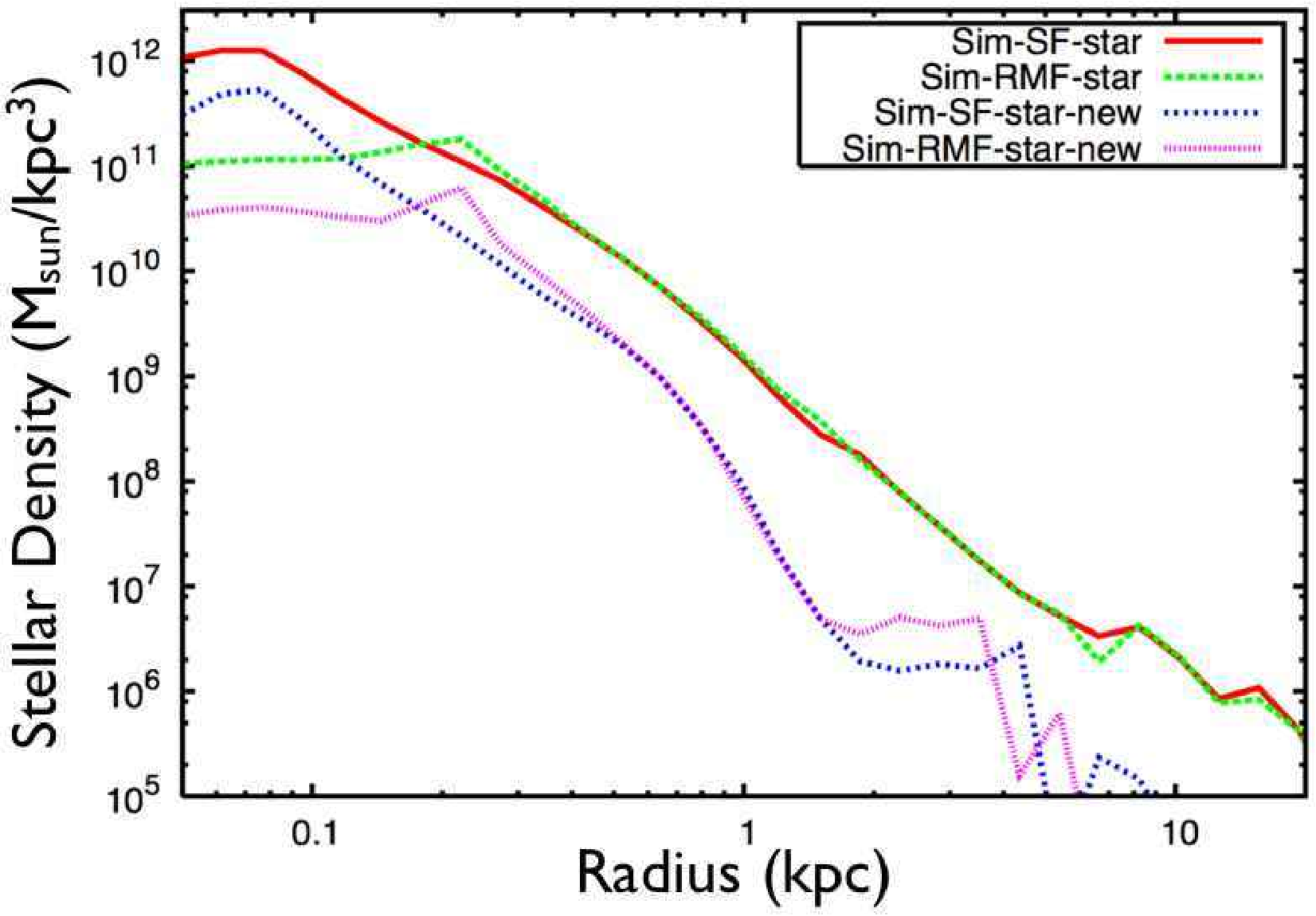}
\caption{Stellar density profiles of total stars and young stars (molecular cloud particles of age $<$ 100 Myr) in a 20 kpc sphere centered on the MBH at $z=2.75$.  The red solid line and the green dashed line represent Sim-SF and Sim-RMF, respectively.  Locally suppressed star formation at the galactic center in Sim-RMF leads to a considerable reduction of stellar density in the region.  
\label{fig:prof_dSdr}}
\end{figure}

The inner core of the galaxy in Sim-RMF becomes a sterile environment for molecular cloud formation (star formation) because the gas is hot and turbulent, therefore Toomre stable.  
As a consequence, star formation is suppressed locally in the inner core, as shown in Figure \ref{fig:prof_dMdr}.
Figure \ref{fig:prof_dSdr} displays how the stellar mass density profile changes in the inner core region as a result of the locally suppressed star formation in Sim-RMF.  
Again we use the snapshot at $z=2.75$, about 220 Myr after the MBH is placed at the center of the model galaxy.
The stellar mass density at 0.1 kpc in Sim-RMF is only less than $\sim$ 20\% of that of Sim-SF.  
The mass density of young stars (age $<$ 100 Myr) shows the similar drastic reduction.  

\begin{figure}[t]
\epsscale{1.18}
\plotone{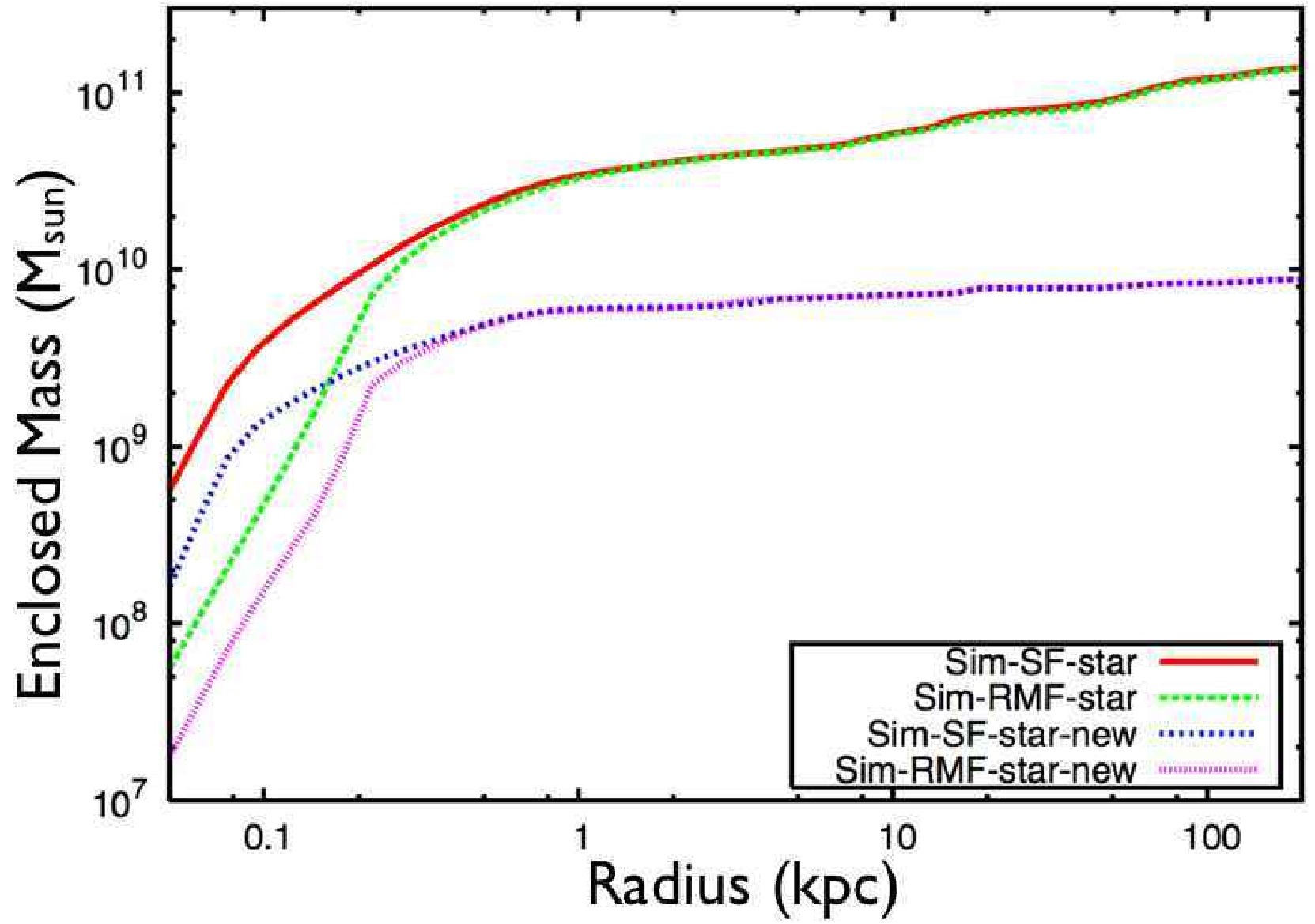}
\caption{Enclosed mass profiles of total stars and young stars (molecular cloud particles of age $<$ 100 Myr) centered on the MBH at $z=2.75$.  The red solid line and the green dashed line represent Sim-SF and Sim-RMF, respectively.  The stellar mass enclosed in the 0.1 kpc sphere of Sim-RMF is about an order of magnitude smaller than that of Sim-SF. 
\label{fig:enc_mass_star}}
\end{figure}

The stellar masses enclosed within a radius r, $M_{\rm star}(<{\rm r})$, are shown in Figure \ref{fig:enc_mass_star}.  
The stellar mass enclosed within the 0.1 kpc sphere of Sim-RMF is almost an order of magnitude smaller than that of Sim-SF. 
The suppressed star formation replaces the steep inner cusp of stellar density profile with a flattened core $\sim$ 0.3 kpc in radius; i.e. stars are less concentrated at the galactic center of Sim-RMF. 
Considering the relatively small amount of energy released from the slowly growing MBH, the difference between these two lines is quite remarkable.
Together with Figure \ref{fig:enc_mass_gas}, one expects that the stellar to gas mass ratio inside the 0.2 kpc sphere of Sim-RMF will be much smaller than that of Sim-SF.  
Note also that the total stellar mass enclosed at the virial radius (80 kpc) is almost indistinguishable between Sim-SF and Sim-RMF.  
In other words, star formation in Sim-RMF is not globally suppressed, but only locally suppressed at the center.
This is because our MBH feedback is not strong enough to unbind a large amount of gas (See \S \ref{sec:4-gas}), or to globally abolish cold, star-forming clumps in the entire ISM.

\begin{figure}[t]
\epsscale{1.18}
\plotone{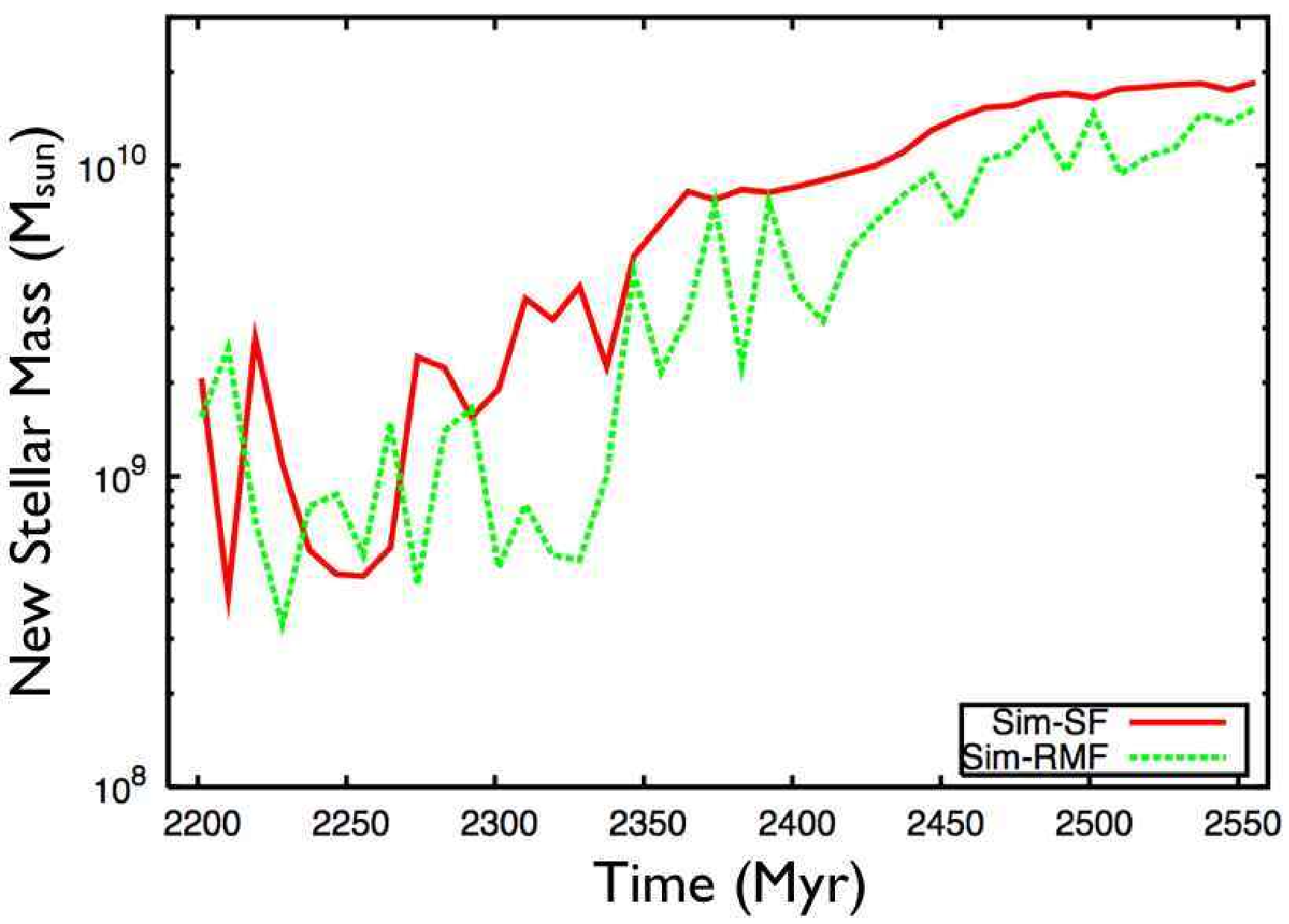}
\caption{Star formation history in a 0.2 kpc sphere: the mass of new stars (molecular cloud particles born after the MBH is placed at 2190 Myr) inside a 0.2 kpc sphere centered on the MBH.  The red solid line and the green dashed line represent Sim-SF and Sim-RMF, respectively.  Since 2300 Myr the star formation activity of Sim-RMF in this region is suppressed.  
\label{fig:SFR}}
\end{figure}

\begin{figure*}[t]
\epsscale{1.17}
\plotone{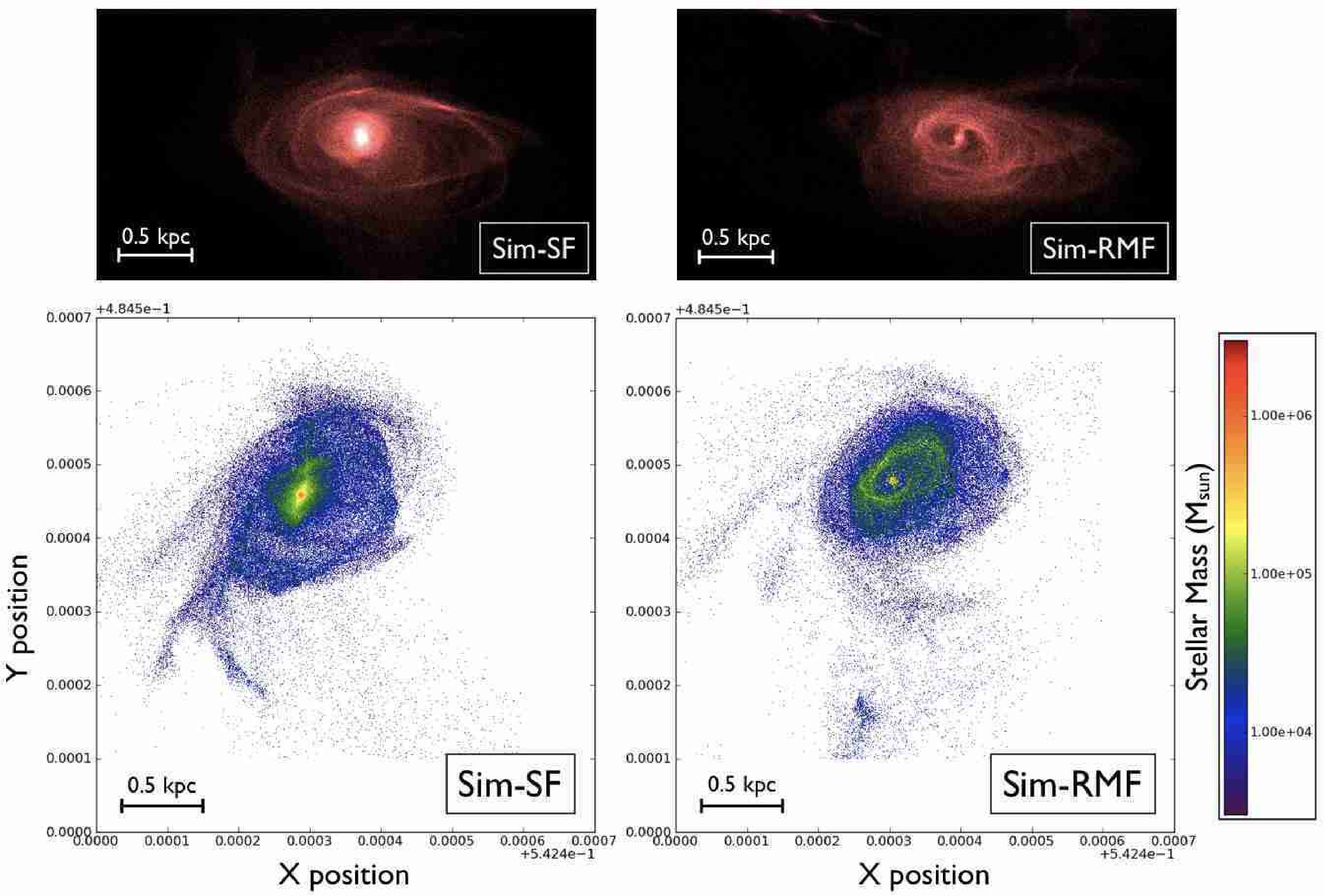}
\caption{The distribution of newly-formed stars (molecular cloud particles of age $<$ 10 Myr) at $z=2.75$.  Sim-SF on the left, and Sim-RMF on the right.  Top: images of newly-formed stars constructed at a $\sim45^{\circ}$ angle from the disk plane.   Visualization courtesy of Ralf Kaehler.  Bottom: newly-formed stellar mass projected along the z-axis of the simulation box which makes a $\sim47.2^{\circ}$ angle with the gas angular momentum vector.  Each frame is 3.0 kpc wide. The difference in stellar distribution is dramatic, especially in the $<$ 0.5 kpc core.  
\label{fig:10Myr}}
\end{figure*}

Figure \ref{fig:SFR} exhibits the evolution of the new stellar mass (molecular cloud particles born after the MBH is placed at 2190 Myr, or at $z=3$)  inside a 0.2 kpc sphere centered on the MBH.  
The plot demonstrates that since 2300 Myr the star formation activity in the inner core of Sim-RMF is suppressed.  
Naturally, alteration in the stellar distribution ensues at the center of the galaxy with an active MBH.
Figure \ref{fig:10Myr} strikingly contrasts the distribution of newly-formed stars (molecular cloud particles of age $<$ 10 Myr).   
\begin{itemize}
\item In the top row, a three dimensional rendering of newly-formed particles is constructed at a $\sim45^{\circ}$ angle from the disk plane.
Here the light intensities of newly-formed particles  are integrated along the lines of sight. 
At the center of the stellar distribution, i.e. the densest peak, lies the MBH particle.  
\item In the bottom row, newly-formed stellar masses are projected along the z-axis of the simulation box which makes a $\sim47.2^{\circ}$ angle with the angular momentum vector of the gas within a 5 kpc sphere centered on the MBH.  
The morphological difference at the inner core of the galaxy is particularly evident.
Stars in Sim-SF are highly concentrated at the center, while stars in Sim-RMF are less concentrated but form spiral-like structures at $\sim$ 0.2 kpc radius from the MBH.  
\end{itemize}

In summary, it is shown that MBH feedback suppresses star formation locally at the galactic inner core, thus significantly changing the stellar distribution there.  
This new channel of feedback is particularly interesting because it is dominant only in the local surroundings of the MBH.
Unlike stellar feedback, which operates globally, this new suppression mechanism does not require additional star formation and/or extensive mass expulsion out of the galactic potential.  

\subsection{Regulated Black Hole Growth}\label{sec:4-MBH}

\begin{figure}[t]
\epsscale{1.12}
\plotone{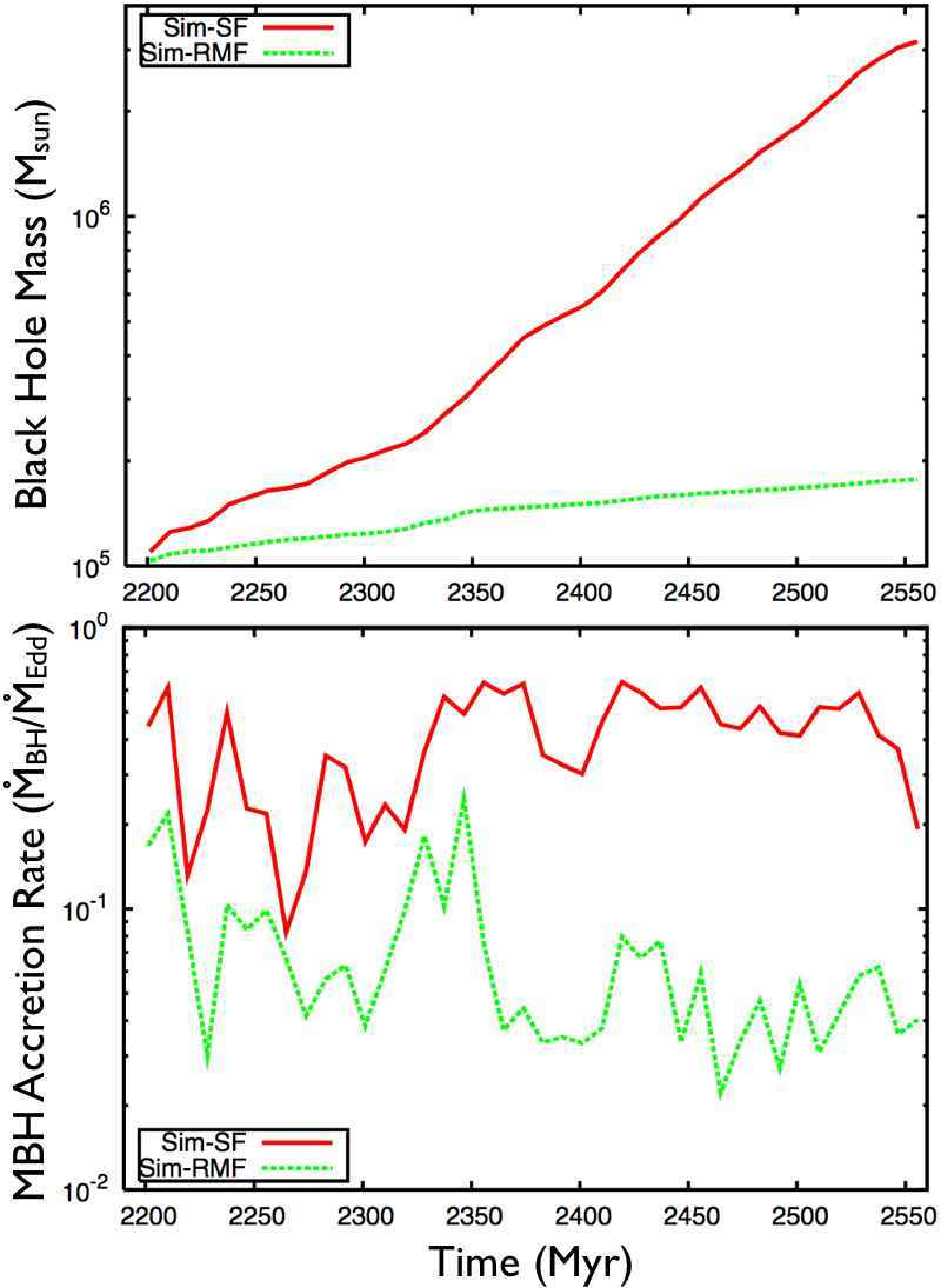}
\caption{Top: black hole mass accretion history.  Note that the mass of the MBH of Sim-RMF has not doubled during this period, while the MBH of Sim-SF has grown exponentially.  Bottom: mass accretion rate onto the MBH in the unit of Eddington rate.  In each panel, the red solid line and the green dashed line represent Sim-SF and Sim-RMF, respectively.  
\label{fig:BHAR}}
\end{figure}

Heating by MBH feedback, which locally suppresses star formation, also makes the MBH to self-regulate its own growth.  Figure \ref{fig:BHAR} shows the MBH accretion history versus time for Sim-SF and Sim-RMF.

\begin{itemize}
\item The MBH in Sim-SF has grown exponentially to $3 \times 10^6 M_{\odot}$ in 350 Myr, already about $>$30 times more massive than its initial mass of $10^5 M_{\odot}$.  
The MBH maintained the accretion rate of 0.2 - 0.6 $\dot M_{\rm Edd}$ during this time period,  corresponding to the unhindered growth of the MBH when there is no mechanism to self-regulate itself other than stellar feedback. 
\item Over the course of the same period the MBH in Sim-RMF has grown by only $\sim$ 70\%  to $1.7 \times 10^5 M_{\odot}$. 
The heated and diffused ISM in the vicinity of the MBH considerably suppresses the Bondi-Hoyle accretion estimate to as low as $\sim 0.02 \,\, \dot M_{\rm Edd}$ in this period. 
This indicates that the MBH feedback described in previous sections is a possible ``radio-mode'' analogue where the accretion rate is $\sim 0.05 \,\dot M_{\rm Edd}$ \citep{2006MNRAS.365...11C}. 
It also presents a potential route to the relatively low mass MBH at the center of the Milky Way ($3 \times 10^6 M_{\odot}$).\footnotemark 
\end{itemize}
\footnotetext{Authors again caution that the initial MBH mass ($10^5 M_{\odot}$) and final masses in both Sim-SF and Sim-RMF lie below the \cite{1998AJ....115.2285M} relationship, when 10\% of the stellar mass is assumed to be in the bulge.
The choice of a relatively small initial MBH mass may have caused a ``radio-mode''-like MBH feedback, and a slower mass growth of the black hole.}

Therefore, the feedback from a MBH is confirmed as an effective mechanism for slowing down the accretion of gas onto itself.  
Without having to suddenly unbind all the surrounding gas, the MBH self-regulates its growth by heating up the neighborhood and keeping it hot for an extended period of time.  
This finding is consistent with the work by \cite{2011MNRAS.412.1341D}, who claimed that the growth of a MBH could be ``self-regulated'', rather than ``supply-limited'' \citep[as in ][]{2005MNRAS.361..776S, 2010arXiv1003.4744T} where quasar-like MBH feedback drive energetic large-scale outflows to unbind a significant amount of gas.

\section{Discussion} \label{sec:5}

\subsection{Summary and Conclusions}\label{sec:5-summary}

A state-of-the-art numerical framework which fully incorporates gas, stars, and a central massive black hole is developed. 
Our simulation, for the first time, followed the comprehensive evolution of a massive star-forming galaxy with self-consistently modeled stars and a MBH. 
Our novel framework renders a completely different, yet physically more accurate picture of how a galaxy and its embedded MBH evolve under each other's influence, providing a powerful means in understanding the coevolution of galaxies and MBHs. 
Our main results and new advancements are as follows.

\begin{enumerate}
\item {\it Molecular Cloud Formation and Feedback:}  
We have included a new model of molecular cloud formation and stellar feedback in our code (\S \ref{sec:2-formation} to \ref{sec:2-SF}).  
Unlike previous star formation recipes based on the Schmidt relation, a particle spawns when a gas cell of a typical molecular cloud size, 15.2 pc,  actually becomes Jeans unstable. 
Then the molecular cloud particle gradually produces stellar mass while returning a large fraction of mass back to the gas with thermal feedback energy, modeling the observed slow star formation in molecular clouds.
Thermal stellar feedback is shown to self-regulates star formation (\S \ref{sec:4-Schmidt}).
\item {\it Massive Black Hole Accretion and Feedback:}
We have successfully developed a self-consistent model of accretion of gas onto a MBH and its radiative and mechanical feedback effects (\S \ref{sec:2-accretion} to \ref{sec:2-MF}).  
Gas accretion onto the MBH is estimated with the Bondi-Hoyle formula, but without any boost factor, as we begin to resolve the Bondi radius. 
Monochromatic X-ray photons from the MBH are followed through three dimensional adaptive ray tracing, rendering the radiative feedback of a MBH; here, rays of photons ionize and heat the gas, and exert momentum onto the gas. 
Finally, the mechanical feedback of the MBH is represented by bipolar jets with velocities of $\sim10^3 \,\,{\rm km \,\,s^{-1}}$ launched from the vicinity of the MBH accretion disk, well resolved in our high-resolution AMR simulations.
Our approach is significantly different from the previous recipes for MBH feedback in galactic scale simulations to date \citep[e.g.][]{2005MNRAS.361..776S, 2007MNRAS.380..877S, 2009MNRAS.398...53B, 2010arXiv1003.4744T, 2011MNRAS.412.1341D}, yet more accurately presents the physics of MBHs when properly incorporated with a high dynamic range.  
\item {\it Locally Suppressed Star Formation:}
By investigating the coevolution of a $9.2\times10^{11} M_{\odot}$ galactic halo and its $10^{5} M_{\odot}$ embedded MBH at $z\sim3$, we show that MBH feedback, especially its radiation, heats the surrounding ISM up to $10^6$ K through photoionization and Compton heating and thus locally suppresses star formation in the inner core of a galaxy (\S \ref{sec:4-gas}).  
The feedback also considerably changes the stellar distribution at the galactic center.
This new channel of feedback from a slowly growing MBH is particularly interesting because it is only locally dominant, and does not require the heating of gas globally on the disk, or instigate a massive gas expulsion out of the galactic potential (\S \ref{sec:4-star}).  
\item {\it Self-regulated Black Hole Growth:}
MBH feedback is also demonstrated to be an effective mechanism for slowing down the accretion of gas onto the MBH itself.  
Without necessarily unbinding all of its surrounding gas, the MBH self-regulates its growth by keeping the surrounding ISM hot for an extended period of time (\S \ref{sec:4-MBH}).  
Therefore, our results possibly are consistent with a ``radio-mode'' analogue of MBH feedback.     
\end{enumerate}

Our method limits the use of {\it ad hoc} formulation and instead more accurately models the physics of galaxy formation.  
As a result, four key components of galactic scale physics, (a) molecular cloud formation, (b) stellar feedback, (c) MBH accretion, and (d) MBH feedback, work self-consistently in one comprehensive framework.  
As an example, the radiation and jets from the MBH heat up the surrounding gas and create hot regions, but the thermal couplings of the radiative and mechanical energy are all carried out by the shock-capturing radiation hydrodynamics AMR scheme itself, not by any presupposed thermal deposition model.  
In our framework, one should also be able to couple small-scale physics (such as molecular cloud formation and feedback) with large-scale physics (such as quasar-driven galactic outflows) without any sub-resolution model.  
These first results undoubtedly demonstrate that we can now develop an unabridged, self-consistent numerical framework for both galaxies and MBHs.

\subsection{Future Work}\label{sec:5-future}

While proven to be fruitful already in producing robust results, our comprehensive galaxy formation framework is only the first step forward in the right direction. 
Imminent future projects and improvements are as follows. 

\begin{enumerate}
\item {\it Merging Galaxies:}  
What is experimented in this work is a quiescent form of MBH feedback, possibly a ``radio-mode'' analogue.  
Meanwhile, MBH feedback is known to make a dramatic difference in galaxy mergers, which is frequent in hierarchical structure formation.  
The gas funneled into galactic centers triggers ``quasar-mode" MBH feedback \citep{2008ApJS..175..356H}, which subsequently reshapes the relation between black hole masses and bulge stellar distributions \citep{2005Natur.433..604D, 2009ApJ...690..802J}.  
Detailed research on this more intense mode of MBH feedback will be the topic of a future paper. 
\item {\it Parameter Studies:} 
More comprehensive parameter studies should follow, especially in the parametrization of MBH feedback and the efficiency of stellar feedback.  
The results should be compared and calibrated with observations such as bulge to disk mass ratio and gas to stellar mass ratio \citep[e.g.][]{2008A&A...482...43G}, or with analytical investigations \citep[e.g.][]{2005MNRAS.358..168S}.  
In particular, the disk-bulge decomposition of simulated galaxies will be the subject of subsequent analysis of our simulations.
\item {\it Improving Mechanical Feedback:}  
In the results presented herein, mechanical feedback is energetically secondary to radiative feedback because the mass accreted onto the MBH is not large enough to repeatedly drive jets.  
These infrequent jets easily penetrate the ISM without necessarily creating sizable shocks or entraining a large amount of gas.
However, a few mechanisms will be considered in the future which could have enhanced the effectiveness of jets. 
Magnetic fields could aid the jets in efficiently depositing outflow momentum onto the infalling gas, as was shown by the studies on the evolution of  jets in the presence of magnetic fields \citep{2009MNRAS.399L..49D, 2010ApJ...709...27W}.  
Cosmic rays accelerated by relativistic jets and shock fronts \citep{2008ApJ...689.1063S} could boost the effectiveness of jets, too.  
\item {\it Improving Radiative Feedback:}  
For now, monochromatic X-ray photons are utilized to carry the energy of MBH radiation (\S \ref{sec:2-RF}); however, a better model will be needed to describe the polychromatic energy distribution of MBH radiation.
Ideally one wants to have a large number of spectral energy bins, each of which is separately followed through three-dimensional ray tracing.
Given the computationally challenging nature of polychromatic radiative transfer, however, tabulated rates of photoionization and photoheating as functions of optical depth can be a good alternative. 
Moreover, to accurately quantify the radiative feedback on the gas in the vicinity of a MBH, the pressure force on dust grains needs to be computed. 
This could have increased the radiation pressure in the presented results, especially in the central $<$ kpc region. 
For this purpose, dust models in \cite{2008MNRAS.383.1281R} will need to be considered. 
\item {\it Adding Supplementary Feedback Channels:}  
A MBH particle in our work represents not just the black hole itself, but also includes accreting gas and stars deep within the galactic nucleus. 
Thus, there is a need for other feedback channels, such as stellar winds from a nuclear disk \citep{2009ApJ...699...89C}. 
The nuclear disk winds can be implemented as thermal deposition of energy, working in conjunction with the aforementioned radiative and mechanical feedback.  
Stellar UV radiation from the nuclear disk can also be incorporated into the radiative feedback of the MBH.  
Including this supplementary feedback will reveal the multi-faceted nature of the coupling of MBH energy with its surroundings. 
\item {\it Improving Accretion Estimate:} 
The accretion estimate using the Bondi-Hoyle formula will need to be improved, especially when the gas disk around the MBH can be resolved down to the Bondi radius. 
Different estimates such as the ones considering gas angular momentum \citep{2010MNRAS.tmp.1085H, 2010ApJ...716.1386L} are attractive candidates that should be explored. 
\item {\it Nonthermal Pressure Sources:}
Nonthermal pressure sources such as magnetic fields \citep{2009ApJ...696...96W}, stellar UV radiation, and cosmic rays are missing in this work, but should be included in future simulations.  
\end{enumerate}

We also recognize that the results from our experiment can provide the community with better sub-resolution models for MBH physics.  
For example, the radial profile of heating rates by the MBH in our simulation can be tabulated; in a coarsely resolved particle-based simulation, one can deposit thermal energy according to this radial dependence into a volume larger than a typical smoothing kernel.  
This can be a useful means for improving the particle-based simulations as well as for speeding up future large-scale AMR calculations, such as the formation of high-redshift quasars and the reionization of intergalactic helium.

\vspace{2 mm}

\acknowledgments

J.K. thanks Steve Allen, T. J. Cox, Nick Gnedin, Oliver Hahn, Patrik Jonsson, Ralf Klessen, Andrey Kravtsov, Yuexing Li, Chung-pei Ma, Michael Norman, Greg Novak, Brian O'Shea, Jeremiah Ostriker, Joel Primack, Romain Teyssier, Matthew Turk, Peng Wang, Risa Wechsler, and an anonymous referee for providing insightful comments and valuable advice.    
J.K. was supported by William R. and Sara Hart Kimball Stanford Graduate Fellowship.  
J.H.W. is supported by NASA through Hubble Fellowship grant \#120-6370 awarded by the Space Telescope Science Institute, which is operated by the Association of Universities for Research in Astronomy, Inc., for NASA, under contract NAS 5-26555.
T.A. acknowledges financial support from the {\em Baden-W\"{u}rttemberg-Stiftung} under grant P-LS-SPII/18, the Heidelberg Institut f\"ur Theoretische Studien. 
J.K. is grateful to Matthew Turk for providing an {\it Enzo} analysis toolkit {\it YT} \citep{2010arXiv1011.3514T}.  
J.K. is indebted to Ralf Kaehler for rendering many beautiful images in this article.    
We gratefully acknowledge the support from Stuart Marshall, Ken Zhou, and SLAC computational team.

\begin{appendix}

\section{A. Accretion Rate Estimate} \label{sec:appendix-A}

The Bondi-Hoyle accretion rate estimate at the molecular cloud formation threshold $n = n_{\rm thres} = 125\, {\rm cm^{-3}}$ and $c_{\rm s} = 10 \,\,{\rm km\,s^{-1}}$ is
\begin{eqnarray}
\dot M_{\rm B} \simeq 0.004 \,\,(M_{\rm BH}/10^5 M_{\odot})^2 \,\,M_{\odot}\, {\rm yr^{-1}}, 
\end{eqnarray}
which is bigger than the Eddington rate, 
\begin{eqnarray}
\dot M_{\rm Edd} \simeq 0.002  \,\,(M_{\rm BH}/10^5 M_{\odot})\,\,M_{\odot}\, {\rm yr^{-1}}.  
\end{eqnarray}
Therefore the density threshold for molecular cloud formation does not limit the accretion rate at any time.  
To put it in another way, because the Bondi-Hoyle rate can surpass the Eddington rate in dense clumps of $n \sim n_{\rm thres}$, the Eddington limit should play a crucial role in restricting the accretion.

We note that, even without the aid of the boost factor (unlike in Eq.(\ref{eq:Springel})), the Bondi-Hoyle estimate in our simulations can surpass the Eddington limit, and averages at 0.2 - 0.6 $\dot M_{\rm Edd}$ in the reported simulation (See \S \ref{sec:4-MBH}). Ê
In other words, had we used the boost factor the Bondi-Hoyle estimate would have been almost always limited by the Eddington limit.  
It is partly because the gas density in our simulations reaches up to $n = n_{\rm thres}$ in the finest cells.  
This justifies our choice of not employing the boost factor which has been common in the previous work \citep[e.g.][and many others]{2005MNRAS.361..776S}.
However, the omission of the usual boost factor does {\it not} indicate that our calculation captures the turbulent accreting flow around the MBH accretion disk.
No contemporary galactic scale simulation - including the reported simulation - has ever captured the turbulent interstellar medium which exists well below the typical resolution limit. 
Therefore, many other models for the MBH accretion estimate are equally applicable in galactic scale simulations, including the Bondi-Hoyle estimate with a boost factor parametrized by the gas density \cite{2009MNRAS.398...53B}.

Readers should keep in mind that the spherical Bondi-Hoyle estimate or any of its variations provides only an approximated accretion rate in a low resolution simulation.  
In reality it is unlikely that the gas around the accreting black hole is spherically symmetric.  
Rather, the gas is expected to reside in a rotationally supported disk which is far from being resolved in a present-day galactic scale simulation.  
In this regard, attempts are being made to improve the Bondi-Hoyle estimate \citep{2010ApJ...716.1386L} or to develop alternatives \citep{2010MNRAS.tmp.1085H, 2010arXiv1003.0605P, 2011MNRAS.412.1341D}.  

\end{appendix}

\end{document}